\documentclass[fleqn,usenatbib]{mnras}

\usepackage{newtxtext,newtxmath}
\usepackage[T1]{fontenc}

\DeclareRobustCommand{\VAN}[3]{#2}
\let\VANthebibliography\thebibliography
\def\thebibliography{\DeclareRobustCommand{\VAN}[3]{##3}\VANthebibliography}

\def\mpcoh{\,h^{-1}{\rm Mpc}} 
\def\gpcoh{\,h^{-1}{\rm Gpc}} 
\def\hompc{\,h\,{\rm Mpc}^{-1}}
\def\msunoh{\,h^{-1}{\rm M}_\odot}
\newcommand{\lyaf}{Ly$\alpha$ forest}
\newcommand{\Lya}{Ly$\alpha$}
\newcommand{\lya}{Ly$\alpha$}
\newcommand{\PoD}{$P_{\rm 1D}$}

\usepackage{graphicx}	% Including figure files
\usepackage{amsmath}	% Advanced maths commands
\title[Lyman alpha mocks with \textsc{AbacusSummit}]{Planting a Lyman alpha forest on \textsc{AbacusSummit}}

\author[B. Hadzhiyska et al.]{Boryana Hadzhiyska,$^{1,2}$\thanks{E-mail: boryanah@berkeley.edu}
A.~Font-Ribera,$^{3}$
A.~Cuceu,$^{4,5,6}$
S.~Chabanier,$^{1}$
J.~Aguilar,$^{1}$
D.~Brooks,$^{7}$
\newauthor
A.~de la Macorra,$^{8}$
P.~Doel,$^{7}$
D.~J.~Eisenstein,$^{9}$
J.~E.~Forero-Romero,$^{10}$
S.~Gontcho A Gontcho,$^{1}$
\newauthor
K.~Honscheid,$^{4,5,6}$
R.~Kehoe,$^{11}$
M.~Landriau,$^{1}$
R.~Miquel,$^{12,3}$
Jundan Nie,$^{13}$
W.J.~Percival,$^{14,15,16}$
\newauthor
G.~Rossi,$^{17}$
Gregory~Tarl\'{e},$^{18}$
and Zhimin~Zhou$^{13}$
\\
$^{1}$ Lawrence Berkeley National Laboratory, 1 Cyclotron Road, Berkeley, CA 94720, USA\\
$^{2}$ University of California, Berkeley, 110 Sproul Hall \#5800 Berkeley, CA 94720, USA\\
$^{3}$ Institut de F\'{i}sica d’Altes Energies (IFAE), The Barcelona Institute of Science and Technology, Campus UAB, 08193 Bellaterra Barcelona, Spain\\
$^{4}$ Center for Cosmology and AstroParticle Physics, The Ohio State University, 191 West Woodruff Avenue, Columbus, OH 43210, USA\\
$^{5}$ Department of Physics, The Ohio State University, 191 West Woodruff Avenue, Columbus, OH 43210, USA\\
$^{6}$ The Ohio State University, Columbus, 43210 OH, USA\\
$^{7}$ Department of Physics \& Astronomy, University College London, Gower Street, London, WC1E 6BT, UK\\
$^{8}$ Instituto de F\'{\i}sica, Universidad Nacional Aut\'{o}noma de M\'{e}xico,  Cd. de M\'{e}xico  C.P. 04510,  M\'{e}xico\\
$^{9}$ Center for Astrophysics $|$ Harvard \& Smithsonian, 60 Garden Street, Cambridge, MA 02138, USA\\
$^{10}$ Departamento de F\'isica, Universidad de los Andes, Cra. 1 No. 18A-10, Edificio Ip, CP 111711, Bogot\'a, Colombia\\
$^{11}$ Department of Physics, Southern Methodist University, 3215 Daniel Avenue, Dallas, TX 75275, USA\\
$^{12}$ Instituci\'{o} Catalana de Recerca i Estudis Avan\c{c}ats, Passeig de Llu\'{\i}s Companys, 23, 08010 Barcelona, Spain\\
$^{13}$ National Astronomical Observatories, Chinese Academy of Sciences, A20 Datun Rd., Chaoyang District, Beijing, 100012, P.R. China\\
$^{14}$ Department of Physics and Astronomy, University of Waterloo, 200 University Ave W, Waterloo, ON N2L 3G1, Canada\\
$^{15}$ Perimeter Institute for Theoretical Physics, 31 Caroline St. North, Waterloo, ON N2L 2Y5, Canada\\
$^{16}$ Waterloo Centre for Astrophysics, University of Waterloo, 200 University Ave W, Waterloo, ON N2L 3G1, Canada\\
$^{17}$ Department of Physics and Astronomy, Sejong University, Seoul, 143-747, Korea\\
$^{18}$ University of Michigan, Ann Arbor, MI 48109, USA\\
}

\date{Accepted XXX. Received YYY; in original form ZZZ}

\pubyear{2015}

\begin{document}
\label{firstpage}
\pagerange{\pageref{firstpage}--\pageref{lastpage}}
\maketitle

\begin{abstract}
The full-shape correlations of the Lyman alpha (Ly$\alpha$) forest contain a wealth of cosmological information through the Alcock-Paczy\'{n}ski effect. However, these measurements are challenging to model without robustly testing and verifying the theoretical framework used for analyzing them. Here, we leverage the accuracy and volume of the $N$-body simulation suite \textsc{AbacusSummit} to generate high-resolution Ly$\alpha$ skewers and quasi-stellar object (QSO) catalogs. One of the main goals of our mocks is to aid in the full-shape Ly$\alpha$ analysis planned by the Dark Energy Spectroscopic Instrument (DESI) team. We provide optical depth skewers for six of the fiducial cosmology base-resolution simulations ($L_{\rm box} = 2\gpcoh$, $N = 6912^3$) at $z = 2.5$. We adopt a simple recipe based on the Fluctuating Gunn-Peterson Approximation (FGPA) for constructing these skewers from the matter density in an $N$-body simulation and calibrate it against the 1D and 3D Ly$\alpha$ power spectra extracted from the hydrodynamical simulation IllustrisTNG (TNG; $L_{\rm box} = 205\mpcoh$, $N = 2500^3$). As an important application, we study the non-linear broadening of the baryon acoustic oscillation (BAO) peak and show the cross-correlation between DESI-like QSOs and our Ly$\alpha$ forest skewers. We find differences on small scales between the Kaiser approximation prediction and our mock measurements of the Ly$\alpha$$\times$QSO cross-correlation, which would be important to account for in upcoming analyses. The \textsc{AbacusSummit} Ly$\alpha$ forest mocks open up the possibility for improved modelling of cross correlations between Ly$\alpha$ and cosmic microwave background (CMB) lensing and Ly$\alpha$ and QSOs, and for forecasts of the 3-point Ly$\alpha$ correlation function. Our catalogues and skewers \href{https://app.globus.org/file-manager?origin_id=9ce29982-eed1-11ed-9bb4-c9bb788c490e&path=%2F}{are publicly available on Globus} via the National Energy Research Scientific Computing Center (NERSC) (full link under Data Availability).
\end{abstract}


%%%%%%%%%%%%%%%%%%%%%%%%%%%%%%%%%%%%%%%%%%%%%%%%%%

%%%%%%%%%%%%%%%%% BODY OF PAPER %%%%%%%%%%%%%%%%%%

\section{Introduction}
\label{sec:intro}
Over the last few decades, we have gained an enormous amount of knowledge about the expansion history of the Universe. With the discovery of the accelerated expansion of the Universe via distance measurements of Type Ia supernovae \citep{Riess:1998AJ....116.1009R,Perlmutter:1999ApJ...517..565P}, an additional ingredient needed to be introduced into the cosmological paradigm. This new component, dubbed ``dark energy,'' took on the responsibility of explaining the mysterious repulsive force these measurements were finding. A couple of decades later, the nature of dark energy is still unknown, and several ongoing and planned surveys have committed to investigating its properties as their top priority (e.g., DESI, DES, Rubin Observatory) \citep{2016arXiv161100036D,2019BAAS...51g..57L,2015AJ....150..150F,2018PhRvD..98d3526A,2016MNRAS.460.1270D,2012arXiv1211.0310L}. %placed as their highest priority the detailed investigation of its properties.%2018PASJ...70S...4A,2021A&A...646A.140H,2015MNRAS.454.3500K,2013LRR....16....6A,2019PASJ...71...43H

These surveys aim to measure the baryon acoustic oscillations (BAO)~\citep{Peebles:1970ApJ...162..815P}, a fixed-scale imprint on large-scale structure that allows us to measure both the angular diameter distance and the Hubble parameter across cosmic time and thus map out the expansion rate of the Universe, bringing important insights into the nature of dark energy. Typically, the BAO peak is measured in the clustering of galaxies, which are used as tracers of the matter density, via the two-point correlation function or the power spectrum \citep[see][for first detections in data]{Eisenstein:2005ApJ...633..560E,Cole:2005MNRAS.362..505C}. Of recent interest are also measurements using quasi-stellar objects (QSOs), which offer an invaluable probe of the $z \sim 1.5$ expansion history of the Universe \citep[e.g.][]{Ata:2018MNRAS.473.4773A}.
% galaxies \cite[e.g.][]{Percival:2010MNRAS.401.2148P,Beutler:2011MNRAS.416.3017B,Blake:2011MNRAS.418.1707B,Alam:2017MNRAS.470.2617A}
In addition, when studying the galaxy and quasar clustering, additional information can be extracted from the amplitude of the redshift-space distortions (RSD), which encodes cosmological information in the form of $f\sigma_8$, a quantity sensitive to the growth of structure. The joint analysis of the growth of structure and the expansion rate has the potential to stress-test general relativity and constrain the various components of our cosmic inventory \citep[see e.g.,][]{2016arXiv161100036D}.
%issues with min separation they can model

%IGM
The Lyman-$\alpha$ forest (\lyaf) provides a powerful alternative probe for glimpsing at our Universe's past. Comprised of a series of absorption features in the
spectra of high-redshift quasars, these spectral features trace the density of neutral hydrogen, and thus the dark matter distribution, on scales larger than the Jeans length \citep{Bi:1992A&A...266....1B}. 
%Indeed, analytical models developed during the 1990s showed that the \Lya\   forest absorption closely traces the distribution of dark matter on scales larger than the Jeans length early in the history of the Universe \cite[e.g.][]{Cen:1994ApJ...437L...9C,Petitjean:1995A&A...295L...9P,Miralda-Escudé:1996ApJ...471..582M}. Since the initial pain-staking efforts to measure the \lyaf\ signal \citep{McDonald:2007PhRvD..76f3009M,Slosar:2011JCAP...09..001S}, our data volume has grown immensely: the Baryon Oscillation Spectroscopic Survey (BOSS) of SDSS-III provided the first BAO detection using the \lyaf\ auto-correlation \citep{Busca:2013A&A...552A..96B,Slosar:2013JCAP...04..026S,Kirkby:2013JCAP...03..024K}, whereas the cross-correlation between the \Lya\   forest and QSOs was first measured in BOSS DR9~\cite{Font-Ribera:2013JCAP...05..018F}, with the first detection of BAO coming in DR11~\cite{Font-Ribera:2014JCAP...05..027F.

{Apart from capturing the BAO feature, quasar spectra speckled with \Lya\  absorption features also contain valuable information on small scales, i.e. several megaparsecs, accessible via the one-dimensional flux power spectrum, \PoD\ \citep{Croft1998,Croft1999,McDonald2000,2001ApJ...557..519Z,Gnedin2002,Croft2002,2004MNRAS.354..684V,McDonald2005,McDonald2006,2006MNRAS.365..231V,2017JCAP...06..047Y,2017MNRAS.466.4332I,Chabanier2019}. Measurements of \PoD, in combination with cosmic microwave background (CMB) probes, have the potential to yield tight constraints on fundamental unknowns such as the sum of the neutrino masses, the shape of the primordial power spectrum, and some exotic dark matter models \citep[see e.g.,][]{Phillips2001,Verde2003,Spergel2003,Viel2004b,Seljak2005,Seljak2006,Bird2011,2017PhRvL.119c1302I,2017JCAP...12..013B,2018PhRvD..98h3540M,2019PhRvL.123g1102M,2019MNRAS.482.3227N,2021PhRvL.126g1302R,2021PhRvD.103d3526R}.}%,PD2015,PD2015b,2006PhRvL..97s1303S,2013PhRvD..88d3502V,2017PhRvD..96b3522I,

%, 3 times as many as in the final eBOSS dataset (approximately 270,000)
The ongoing Dark Energy Spectroscopic Instrument (DESI)  survey will achieve an unprecedented precision in the \lyaf\ measurements across all scales, amassing approximately a million quasar spectra at $z > 2$ over its five years of operation \citep[for various specifications on the experiment, see][]{2013arXiv1308.0847L,2016arXiv161100037D,2022arXiv220510939A,2022arXiv220509014S,2023ApJ...944..107C}. 
Ahead of such immense improvements in our statistics, a factor of four larger than current surveys, it is crucial that we diligently stress test our analysis pipelines and quantify the impact of secondary astrophysical effects. The most viable path forward is through the development of synthetic mock datasets \cite[e.g.][]{LeGoff:2011A&A...534A.135L,Font-Ribera:2012JCAP...01..001F,Bautista:2015JCAP...05..060B,2014ApJ...784...11P,2022MNRAS.514.3222P,2016ApJ...827...97S,2020JCAP...03..068F,2022ApJ...927..230S}, which must strike the careful balance of computational efficiency and survey realism.

In this work, we provide a new mock dataset, which aims to build upon previous such efforts in several key ways. Other large-scale mocks adopted in the literature tend to compromise on the accuracy of their \lyaf\ model, for example, by utilizing lognormal realizations instead of dark matter simulations, placing a greater emphasis on volume. {Our mocks, on the other hand, are generated using the $N$-body simulation suite, \textsc{AbacusSummit}, and therefore provide greater realism in the non-linear regime than the lognormal mocks while also covering a sufficient volume of $\sim$100 Gpc$^3$ to satisfy the requirements of the DESI survey. In addition, the model used to create them is calibrated on the state-of-the-art hydrodynamical simulation IllustrisTNG and thus has an advantage over standard approaches for modeling the large-scale \lyaf\ signal.} At the same time, it is simple enough that it can be applied to an arbitrarily large number of simulations, without this exercise becoming prohibitively expensive, as in the case of the hydrodynamical simulations used in \PoD\ analysis. 

A second major goal of this work is to integrate the 1D and 3D correlation function analyses. Typically, the BAO and \PoD\ analyses are carried out as independent probes, with the BAO measurements being modeled via linear perturbation theory, while the \PoD\ ones via hydrodynamical simulations that capture the physics of the intergalactic medium (IGM). The joint analysis of these measurements would not only improve the statistical uncertainty on
cosmological parameters, but also make them more robust to systematic errors \citep{2018JCAP...01..003F}. In order to accomplish this, however, we need a theoretical
framework that can be trusted on all scales. While the mocks presented in this work lack the gas and IGM physics needed to reliably model the smallest scales targeted by \PoD\ analyses, $k \sim 10 \hompc$, they still support cosmological scales spanning several orders of magnitude, $0.001 \lesssim k \lesssim 1 \hompc$. They thus allow an excellent opportunity to develop novel pipelines and statistics, beyond the standard BAO analysis, for extracting cosmological information from the full shape of the 3D correlations \citep[see e.g.,][]{2021MNRAS.506.5439C}. Such work is planned by the DESI collaboration in the near term, and our mocks provide an important first step towards reaching these goals. {As an example, these mocks provide realistic connection between the QSOs and the \lyaf, allowing for accurate modeling of their cross-correlation down to intermediate and small scales, which typically elude more simplistic mocks. Given the high resolution and large volume of the \textsc{AbacusSummit} simulations, the mocks presented in this work can be used to develop high-fidelity models for analyzing upcoming measurements of the \lyaf.}
%In subsequent work, we plan to develop new techniques that allow us to improve the accuracy and realism of our mocks such as phase-space tessellation for depositing the dark matter particles onto a grid \citep{2012MNRAS.427...61A}, light cones for obtaining  properly redshift-evolved spectra in the curved sky, and machine-learning inspired approaches for ``learning'' the dark-matter-flux relationship from hydro simulations.

This paper is organized as follows. In Section~\ref{sec:meth}, we introduce the simulations and summary statistics employed in this study. In Section~\ref{sec:mocks}, we detail our procedure for generating the \lyaf\ mocks and present a comparison with the high-vericity \Lya\  skewers extracted from the hydrodynamical simulation IllustrisTNG. In Section~\ref{sec:stats}, we show the outcome of applying our algorithm to six of the $N$-body simulation suite boxes of \textsc{AbacusSummit}. In particular, we examine the 1D and 3D power spectra as well as the auto- and cross-correlation of the \lyaf\ and QSOs, demonstrating the impact of non-linear clustering on these observables. We summarize our findings and discuss relevant implications about future work in Section~\ref{sec:conc}.

\section{Methods}
\label{sec:meth}

\subsection{Simulations}
\label{sec:sim}
In this Section, we introduce the two simulation suites relevant to this work: IllustrisTNG and \textsc{AbacusSummit}.

\begin{figure}
    \centering
    \includegraphics[width=0.48\textwidth]{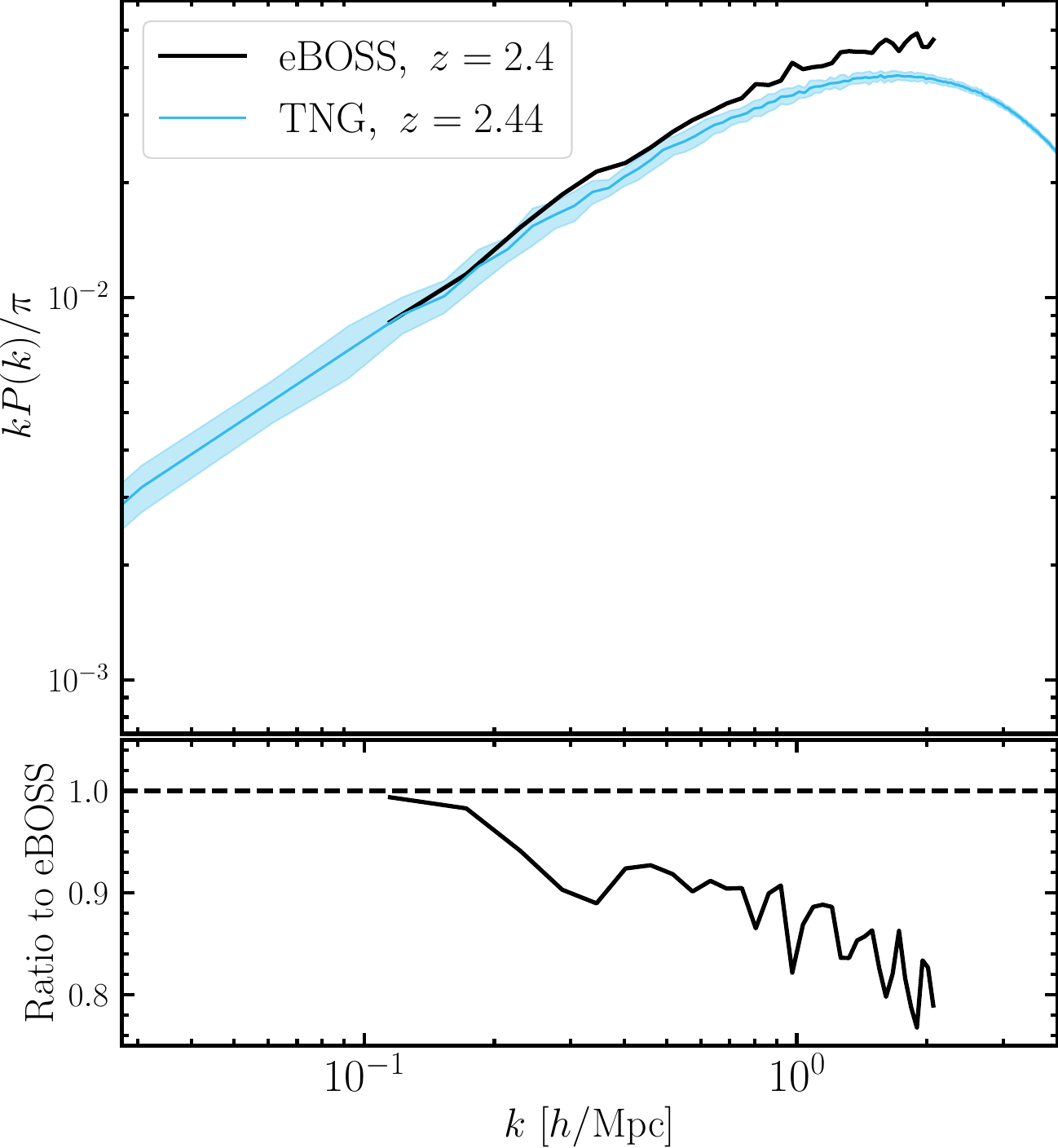}\\
    \includegraphics[width=0.48\textwidth]{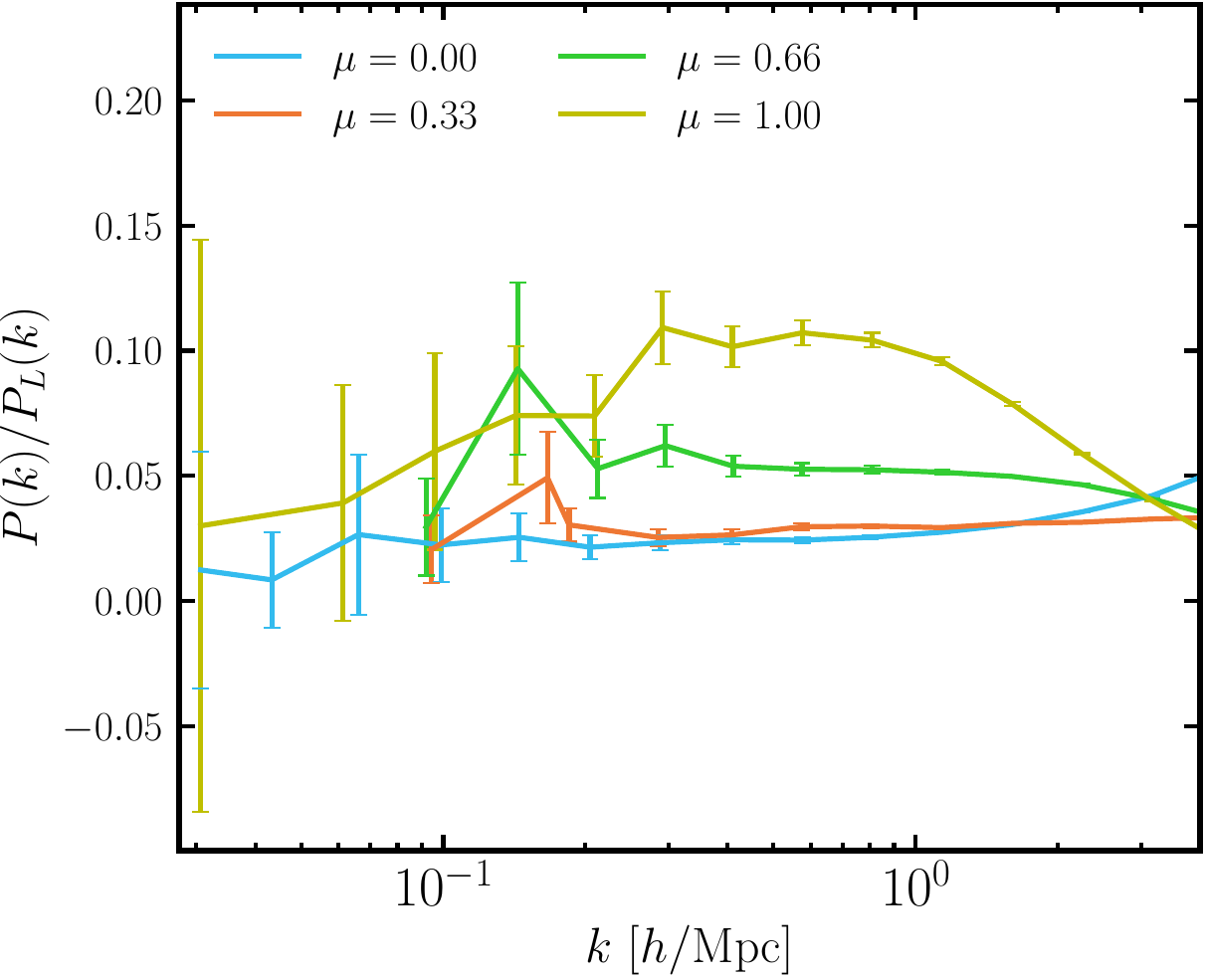}
    \caption{
    \textit{Top panel:} Comparison of the 1D power spectrum of the \lyaf\ skewers extracted from TNG300-1 (in blue) and eBOSS DR14 (in black) \citep{Chabanier2019}. We note that the eBOSS power spectrum is already noise and background substracted, and the Ly$\alpha$-Si correlations have been removed, facilitating the comparison with the simulation. The lower segment of the plot shows the ratio with respect to eBOSS. The agreement between simulations and observations is reasonably good. In particular, the discrepancy is $\lesssim 10\%$ for $k \lesssim 1 \hompc$, and then increases up to 20\% on smaller scales, more prone to resolution effects. \textit{Bottom panel:} 3D power spectrum of the TNG300-1 \lyaf\ skewers, with values of the bias and the redshift distortion parameter given by $b_{{\rm Ly\alpha, TNG}} = -0.1379$ and $\beta_{{\rm Ly\alpha, TNG}} = 1.432$, respectively. These are slightly different from the eBOSS constraints of $b_{{\rm Ly\alpha, eBOSS}} = -0.117$ and $\beta_{{\rm Ly\alpha, eBOSS}} = 1.669$ at $z_{\rm eff} = 2.334$ \citep{2020ApJ...901..153D}, but still comparable and sufficiently close to the observations for the purposes of this study.}
    \label{fig:p1d_eboss}
\end{figure}

\subsubsection{IllustrisTNG}

The Next Generation Illustris simulation (IllustrisTNG, TNG), which is run with the AREPO code \citep{2010MNRAS.401..791S,2019arXiv190904667W}, consists of 9 simulations: 3 box sizes (300, 100 and 50 Mpc on a side), each available at 3 different resolutions, 1--3, with 1 being the highest and 3 the lowest resolution \citep[see][for details]{2018MNRAS.475..676S,2018MNRAS.477.1206N,2018MNRAS.480.5113M,2019MNRAS.tmp.2010N,2019MNRAS.tmp.2024P}. Compared with its predecessor, Illustris \citep{2014MNRAS.444.1518V,2014Natur.509..177V,2014MNRAS.445..175G},  TNG provides improved agreement with observations by modifying its treatment of active galactic nuclei (AGN) feedback,  galactic winds and magnetic fields \citep{2018MNRAS.473.4077P,2017MNRAS.465.3291W}. In addition, various improvements of the hydrodynamical convergence have been introduced in the code. 

In this work, we employ the highest-resolution hydro run of the largest TNG box, TNG300-1, as well as the lowest-resolution dark-matter-only run, TNG300-3-DM, at $z = 2.44$, which we use to calibrate our \lyaf\ generation procedure (see Section~\ref{sec:mocks}). Having a phase-matched dark-matter-only simulation allows for a fast and direct comparison to the full hydro results. In particular, since the sample variance of the two boxes is the same, any differences observed in the power spectra can be attributed to model choices. TNG300-3-DM also has the benefit of having very similar particle resolution to the \texttt{base} \textsc{AbacusSummit} boxes: $M_{\rm part, TNG-300-3-DM} = 3.1 \times 10^9\msunoh$ and $M_{\rm part, Abacus} = 2.1 \times 10^9\msunoh$.

We make use of the noiseless mock \lyaf\ spectra created and made publicly available by \citet{2022ApJ...930..109Q}. The spectra are obtained via the \texttt{fake\_spectra} package \citep{2015MNRAS.447.1834B,2017ascl.soft10012B}, which calculates the absorption spectra for every ion in the simulation along a chosen set of lines-of-sight. Each particle contributes to the overall absorption in the spectrum according to a Voigt profile. The cells are smoothed by an appropriate top-hat kernel. {The skewers used in this study are solely due to Ly$\alpha$ transmission, and we leave further exploration of the effect of metal absorption lines on the Ly$\alpha$ observables for future work.}

Throughout this work, we assume that the mean flux evolution is given by the following empirical relation, corrected for metal absorption \citep{2008ApJ...688...85F}:
\begin{equation}
    \langle F \rangle = {\rm exp} [-1.330 \times 10^{-3} \times (1+z)^{4.094}].
    \label{eq:mean}
\end{equation}
In our \lyaf\ skewers, the optical depth is rescaled to match the expected observed measurement, which at $z = 2.44$ corresponds to $\langle F \rangle = 0.8101$. The high-resolution spectra assume the line-of-sight direction to be along the $z$ axis, with a pixel width of 6.4 ${\rm km}/s$ {to resolve well features along the line-of-sight.}
%capture the effect of saturation. 
We note that ideally one would use all three axes as lines-of-sight to reduce the variance of the measurements. However, those were not provided as part of the TNG \lyaf\ data release. Once the skewers are extracted, the transmission fraction is averaged over adjacent pixels to a final pixel size of 26 ${\rm km}/s$, corresponding to 0.25 $\mpcoh$. When employing its dark-matter-only counterpart, TNG300-3-DM, to generate the mock skewers, we adopt a pixel size of 0.33 $\mpcoh$ (625$^3$ cells), corresponding also to the interparticle spacing of the simulation and approximately matching the resolution of the \textsc{AbacusSummit} boxes (0.29 $\mpcoh$). %Finally, to mimic the spectral broadening, the spectra are smoothed with a Gaussian kernel of $\sigma = 2.06$ \AA \ in the observed frame.

\subsubsection{Comparing TNG300-1 with eBOSS}
\label{sec:tng_eboss}

We next compare the one-dimensional (1D) power spectrum measured from the TNG300-1 simulation with observational data. The observational results presented here are based on data collected by the Sloan Digital Sky Survey (SDSS) \citep{2000AJ....120.1579Y}. In particular, the sample of \lyaf\ forest observations is selected from the quasar spectra of the DR14 catalog, which were observed either by the SDSS-III Collaboration between 2009 and 2014 (as part of BOSS) or by the SDSS-IV Collaboration in 2014-2015 (as part of eBOSS). Throughout the paper, we will be referring to this data set as `eBOSS.'

In Fig.~\ref{fig:p1d_eboss}, we illustrate the 1D power spectrum of the \lyaf\ skewers measured in TNG300-1 and weigh it up against eBOSS DR14 \citep{Chabanier2019}. We note that the eBOSS power spectrum is already noise and background substracted, and the Ly$\alpha$-Si correlations have been removed, allowing for a direct comparison with the simulation. 
{It is important to acknowledge that we do not expect a perfect agreement between observations and simulations, since uncertainties in the thermal and ionisation history of the intergalactic medium impact the correlations of the Ly$\alpha$ forest.}
%It is important to acknowledge that we do not expect a perfect agreement between observations and simulations on small scales, as small-scale observables such as the 1D power spectrum are strongly affected by astrophysical and thermodynamical processes, implemented as subgrid models in current hydrodynamical simulations, whose impact varies significantly between different simulation codes \citep[e.g., the AGN feedback prescription can lead to $\gtrsim 10\%$ differences in the power spectrum inferred from IllustrisTNG; see][]{2017ApJ...835..175G}. 
We see that the agreement between simulations and observations is reasonably good, noting that 
%our TNG measurements are extracted from a relatively low-resolution simulation grid along the line-of-sight (820 pixels) and that 
we do not calibrate the mean flux of TNG to the observed mean in eBOSS data, which would change the overall normalization. We observe a discrepancy of $\lesssim 10\%$ for $k \lesssim 1 \hompc$, and up to 20\% on smaller scales. We have checked that a finer gridding of the TNG gas cells along the line-of-sight (50 ckpc$/h$) does not mitigate the small-scale deviation. For the purposes of this work, this is a satisfactory result. However, we note that TNG has not been exhaustively tested against observations in the IGM regime (but rather mostly for galaxy observations) unlike other hydro simulations tailored towards mimicking the \lyaf\ \citep[see e.g., the Nyx vs. Illustris code comparison in][]{2018ApJ...859..125S}. It is also worth commenting on the fact that our result exhibits greater tension with eBOSS than can be seen in the analogous figure in \citet{2022ApJ...930..109Q}. The reason for this difference is that the eBOSS data vector of \citet{2022ApJ...930..109Q} uses a different technique for subtracting the noise contributions than the official eBOSS analysis\footnote{Established through private communication.} \citep{Chabanier2019}.

In the lower panel of Fig.~\ref{fig:p1d_eboss}, we show the 3D power spectrum defined as follows:
\begin{equation}
        \label{eq:pkmu}
        \langle \tilde{\delta_F}(\mathbf{k}) \tilde{\delta_F}^*(\mathbf{k}') \rangle = (2 \pi)^3 P(k, \mu) \delta_D(\mathbf{k} - \mathbf{k}') ,
\end{equation}
where $\delta_D(\mathbf{k})$ is the three-dimensional Dirac delta function. In particular, we bin the power spectrum $P(k, \mu)$ into 20 logarithmic $k$ bins ranging from $k \in \{(2 \pi)/L_{\rm box}, \ 15\hompc$, {where $L_{\rm box}$ is the box size of the simulation,} and 16 $\mu$ bins ranging from 0 to 1, and show the estimated Gaussian error bars (see further discussions in Section~\ref{sec:tune} and Section~\ref{sec:power}). It is evident that on large scales ($k \lesssim 0.1 \hompc$), the error on the measurements is large due to the small size of the box. We compare the fitted values of the \Lya\ bias and redshift distortion parameter from TNG, $b_{{\rm Ly\alpha, TNG}} = -0.1379$, $\beta_{{\rm Ly\alpha, TNG}} = 1.432$, to the values measured in eBOSS data, $b_{{\rm Ly\alpha, eBOSS}} = -0.117$, $\beta_{{\rm Ly\alpha, eBOSS}} = 1.669$ at $z_{\rm eff} = 2.334$
\citep[see last column of Table 6 in][]{2020ApJ...901..153D}. 
{While the redshift distortion parameter is slightly lower in TNG, it is interesting to see that the values of $b (1+\beta)$ that set the amplitude of the 3D power along the line of sight are within 10\%.}
%We find that both the bias as well as the redshift distortion parameter are slightly lower in TNG. 
We note that other state-of-the-art hydro simulations report higher values for $\beta$ (e.g., Chabanier et al. in prep. find $\beta=1.8$), {while \citet{2022JCAP...09..070G} finds $\beta=1.35$ at $z=2.8$ in the Sherwood suite of simulations and \citet{2015JCAP...12..017A} find $\beta \sim 1.3 -1.5$ in the relevant redshift range}. 
However, for the purposes of this study, we consider these matches good enough and calibrate our mocks to match the TNG measurements, referring to it as the `truth' from hereon.
%bias, beta -0.13785625,  1.43235174
%b(1+beta) TNG=0.335   eBOSS=0.312

\subsubsection{\textsc{AbacusSummit}}

\textsc{AbacusSummit} is a suite of high-performance cosmological $N$-body simulations, which was designed to meet and exceed the Cosmological Simulation Requirements of the DESI survey \citep{2021MNRAS.508.4017M}. The simulations were run with \textsc{Abacus} \citep{2019MNRAS.485.3370G,2021MNRAS.508..575G}, a high-accuracy cosmological $N$-body simulation code, optimized for GPU architectures and  for large-volume simulations, on the Summit supercomputer at the Oak Ridge Leadership Computing Facility. 

The majority of the \textsc{AbacusSummit} simulations are made up of the \texttt{base} resolution boxes, which house 6912$^3$ particles in a $2\gpcoh$ box, each with a mass of $M_{\rm part} = 2.1 \ 10^9\msunoh$. While the \textsc{AbacusSummit} suite spans a wide range of cosmologies, here we focus on the fiducial outputs (\textit{Planck} 2018: $\Omega_b h^2 = 0.02237$, $\Omega_c h^2 = 0.12$, $h = 0.6736$, $10^9 A_s = 2.0830$, $n_s = 0.9649$, $w_0 = -1$, $w_a = 0$). In particular, we employ the 6 \texttt{base} boxes \texttt{AbacusSummit\_base\_c000\_ph\{000-005\}}. The reason for our choice is that full particle outputs are provided for these simulations at $z = 2.5$, which is the redshift of interest for our \lyaf\ study. For full details on all data products, see \citet{2021MNRAS.508.4017M}. In future work, we plan to extend our mocks to cosmologies beyond \textit{Planck} 2018 and adapt our method so that it utilizes only 10\% of the particles (available for all \textsc{AbacusSummit} simulations at $z = 2.5$).

\subsection{Quasar catalogue}
\label{sec:quasar}

The cross-correlation function of the \lyaf\ with quasars will be measured by current and next-generation experiments such as DESI. However, to ensure that our theoretical models can adequately fit the signal, we need to test our pipelines on synthetic catalogs. To this end, we generate mock quasar catalogues via \textsc{AbacusHOD}, a sophisticated routine that builds upon the baseline halo occupation distribution (HOD) model by incorporating various extensions affecting both the one- and two-halo terms, and in Section~\ref{sec:corr}, we show the cross-correlations of our mock quasar catalogue with the \lyaf\ spectra. \textsc{AbacusHOD} allows the user to specify different tracer types: emission-line galaxies (ELGs), luminous red galaxies (LRGs), and quasistellar objects (QSOs). The full model is described in detail in \citet{2022MNRAS.510.3301Y}. 

In this study, we adopt a simple HOD model for the QSO without any decorations:
\begin{align}
    \bar{n}_{\mathrm{cent}}^{\mathrm{QSO}}(M) & = \frac{\mathrm{ic}}{2}\mathrm{erfc} \left[\frac{\log_{10}(M_{\mathrm{cut}}/M)}{\sqrt{2}\sigma}\right], \label{equ:zheng_hod_cent}\\
    \bar{n}_{\mathrm{sat}}^{\mathrm{QSO}}(M) & = \left[\frac{M-\kappa M_{\mathrm{cut}}}{M_1}\right]^{\alpha}\bar{n}_{\mathrm{cent}}^{\mathrm{QSO}}(M),
    \label{equ:zheng_hod_sat}
\end{align}
where $M_{\mathrm{cut}}$ characterizes the minimum halo mass to host a central galaxy, $M_1$ the typical halo mass that hosts one satellite galaxy, $\sigma$ the steepness of the transition from 0 to 1 in the number of central galaxies, $\alpha$ the power law index on the number of satellite galaxies, {${\rm ic}$ the incompleteness parameter}, and $\kappa M_\mathrm{cut}$ gives the minimum halo mass to host a satellite galaxy. The parameters we choose for our QSO catalogs are in units of $\msunoh$
\begin{eqnarray}
    \log_{10}{(M_{\rm cut})} = 13.2, \ \
        \kappa = 1.11, \ \
        \sigma = 0.65, \\
        \log_{10}{(M_1)} = 13.8, \ \ 
        \alpha = 0.8, \ \ {\rm ic} = 1.0, \nonumber
\end{eqnarray}
which have been selected so as to yield a linear bias of about $b^{\rm QSO} \approx 3.3$, roughly matching the quasar bias in \citet{2020ApJ...901..153D}, and have a number density of $1.75\times 10^{-4} \ [\mpcoh]^{-3}$ (i.e., 1.4 million quasars per box). These numbers are taken from rough fits to preliminary DESI data.

%\subsection{Dark Energy Spectroscopic Instrument}

%The main purpose of this work is to provide accurate and realistic mocks of the \lya\ absorption at $z > 2$ that will aid the planned analysis of the \lyaf\ by the Dark Energy Spectroscopic Instrument (DESI) team.

%DESI is a Stage IV dark energy experiment currently conducting a five-year survey of about a third of the sky with the goal to amass spectra for approximately 40 million galaxies and quasars \citep{2016arXiv161100036D}. The instrument operates on the Mayall 4-meter telescope at Kitt Peak National Observatory \citep{2022arXiv220510939A} and can obtain simultaneous spectra of almost 5000 objects over a $\sim$3$^{\circ}$ field \citep{2016arXiv161100037D,2022arXiv220509014S} thanks to a robotic, fiber-fed, highly multiplexed spectroscopic surveyor. The goal of the experiment is to unravel the nature of dark energy through precise measurements of the expansion history of the Universe \citep{2013arXiv1308.0847L} and thus the dark energy equation of state parameters $w_0$ and $w_a$, with a predicted factor of five to ten improvement on their error relative to previous Stage-III experiments \citep{2016arXiv161100036D}.

\section{Creation of the mocks}
\label{sec:mocks}

Previous large-volume \lyaf\ mocks have been generated using simple, fast and computationally cheap methods such as lognormal density maps \citep[e.g.,][]{2020JCAP...03..068F} augmented with approximate prescriptions to reach the volumes required by the new generation of surveys. However, models based on Gaussian random fields do not capture non-linear evolution, as they are generated solely through the initial power spectrum \cite{Coles:1991MNRAS.248....1C,Bi:1997ApJ...479..523B}. Slightly more complex are formalisms involving Lagrangian perturbation theory \citep[see e.g.,][for a review]{Bernardeau:2002PhR...367....1B} and COLA \cite{Tassev:2013JCAP...06..036T}, which extend the modeling capabilities to the mildly non-linear regime. In pure BAO analyses, the presence of non-linear structure does not substantially affect the measurement, especially at the high-redshift regime ($z \gtrsim 2$). However, any full-shape and small-scale analysis of \lyaf\ observables (including cross-correlations) will be substantially impacted by non-linear graviational and astrophysics effects \citep{2022arXiv220912931C,2022arXiv220913942C}. 

This work aims to enable the full-shape analysis of the \lyaf\ power spectrum, planned to be conducted as part of the DESI Y3 \Lya\  science program. While ideally one would strive to generate as realistic mocks as possible, which would mean employing state-of-the-art hydrodynamical simulations, this is unfortunately not a viable path forward, as the computational expense associated with generating \lyaf\ skewers in a volume sufficiently large for modern surveys is tremendous. In this work, we therefore seek a middle path of using fully evolved $N$-body simulations 
%(\textsc{AbacusSummit}) 
and adopting an approximate technique
%, the Fluctuating Gunn-Peterson Approximation (FGPA), 
calibrated to a hydro simulation. 
%IllustrisTNG.

In our \lyaf\ mocks on \textsc{AbacusSummit}, we opt for a resolution of 6912$^3$ cells per box, corresponding to an average of one particle per cell and a mean interparticle distance of 0.29 $\mpcoh$. The density and velocity field grids are obtained as described in Section~\ref{sec:dens}. The resolution is chosen to be comparable to (though still larger than) the Jeans length at that redshift (100 kpc/$h$) while avoiding the creation of too many empty cells, as that would contribute substantial noise to the density field and the derived optical depth, subsequently. Since our resolution is limited by the simulation resolution, we are unable to obtain an accurate estimate of the field at scales lower than $\sim$0.3 $\mpcoh$. Thus, the power spectrum of the skewers $P_{1\mathrm{D}}(k_\parallel)$ measured from modes lying along the line of sight is suppressed, which also affects the 3D flux power spectrum \citep{2020JCAP...03..068F}. 
%This subsequently affects the errors on our BAO measurements, as the 3D flux power spectrum of the \Lya\   forest has a significant contribution to its error that is proportional to the 1D power spectrum, known as aliasing noise~\cite{McDonald:2007PhRvD..76f3009M}. 
For this reason, we boost the power spectrum by adding small-scale fluctuations to the density field, as discussed in Section~\ref{sec:noise}.

Next, to convert from dark matter density to optical depth, we adopt the simple fluctuating Gunn-Peterson approximation (FGPA) \cite{Croft:1998ApJ...495...44C}. While this method is simplistic, it offers a fast and transparent way of connecting the matter density to that of neutral hydrogen. {More complex techniques do exist, including the \lya\ Mass Association Scheme (LyMAS; \citet{Peirani:2014,Peirani:2022}), the Iteratively Matched Statistics (IMS; \citet{Sorini:2016} method, and Hydro-BAM \citep{2022ApJ...927..230S}. These use a variety of approaches tuned using smaller hydro simulations that range from matching the \lyaf\ probability distribution function and/or power spectrum to using a supervised machine learning method. However, these methods have yet to be applied to simulations with the purpose of making large scale DESI mocks. Therefore, in this first work we focus on using the simpler FGPA approach and leave the application of these more complex recipes to future work.}
% We reserve the application of more complex recipes for future work \citep[see e.g.,][]{Irsic:2018JCAP...04..026I}. 
We adopt two slight variations of the FGPA approach discussed in Section~\ref{sec:opt}.

Finally, we add RSDs to our skewers and convert them to transmission flux spectra in Section~\ref{sec:flux}. Those are the result of peculiar velocities in the inter-galactic medium (IGM) projected along the line of sight, and manifest themselves as an anisotropy in the power spectrum and correlation function measurements.

\subsection{Calculating the density and velocity fields}
\label{sec:dens}

The first step in applying the FGPA method to an $N$-body simulation (in our case, \textsc{AbacusSummit} and TNG300-3-DM) is the deposition of particles onto a grid. In this study, we adopt triangular shape cloud (TSC) interpolation, to obtain both the density, $\rho_{\rm dm} (\mathbf{x})$, and peculiar velocity  $v_{r} (\mathbf{x})$ fields. Some studies \citep[e.g.,][]{2016ApJ...827...97S} apply smoothing to the fields to mimic the effect of baryonic pressure on small scales. However, similarly to \citet{2022ApJ...930..109Q,2015MNRAS.453.4311S,2020ApJ...891..147N}, we find that when simulating the neutral hydrogen absorption on scales of $\sim$1 Mpc, smoothing the fields has negligible effects. In future work, we plan to revisit our choice of a particle-to-grid deposition method. While TSC has clear advantages (especially in the low-density regime) over the lower-order kernels, i.e. nearest grid point (NGP) and cloud-in-cell (CIC), tessellation-based methods are even better suited for obtaining a near exact estimate of the low-density dark matter field (e.g., phase-space sheet tesselation as in \citet{2012MNRAS.427...61A}), which is the most relevant for \lyaf\ analysis \citep[see e.g.,][for an evaluation of these effects]{2023MNRAS.518.3754C}. 

\subsection{Adding small-scale noise}
\label{sec:noise}

Similarly to \citet{2020JCAP...03..068F}, we add small-scale noise to the initial density field so as to make up for the deficit in the 1D power spectrum. This deficit is the result of the effective smoothing on small scales imposed by the relatively large size of the gas blobs ($\sim0.3\mpcoh$), which suppresses the power on small scales. We start by generating independent Gaussian skewers $\delta_\epsilon$ for each line-of-sight preserving the resolution of the original density field. We impose that the 1D power spectrum of the skewers obeys the following equation \cite{McDonald2006}:
\begin{equation} 
    \label{eq:noise}
    P_{\rm{1D}}(k)~\propto~[1~+~(k/k_1)^n]^{-1},
\end{equation}
normalized so that ${\rm Var}[\delta_\epsilon] = 1$. These Gaussian skewers are then all scaled by a factor $\sigma_\epsilon$ to control the variance in the extra power added. This factor, together with $n$ and $k_1$ is a free parameter in our model and takes a single value determined by the procedure described in Section~\ref{sec:tune}.

The new density skewers are then obtained by multiplying the true density field skewers by the lognormal field:
\begin{equation} 
    \label{eq:lognorm}
    \rho (\textbf{x}) = \rho_{\rm dm} (\textbf{x}) (1 + \delta_{\rm ln} (\textbf{x})),
\end{equation}
where the lognormal field is given by the lognormal transformation
\begin{equation}
    \delta_{\rm ln} (\textbf{x}) = {\rm exp}\left[\delta_\epsilon (\textbf{x}) \sigma_\epsilon - \frac{\sigma_\epsilon ^2}{2}\right] - 1
\end{equation}
to ensure zero mean.

Note that the lognormal skewers are generated independently of each other, so there is no correlation across different lines-of-sight. 
%For this reason, they only affect the 3D correlations by contributing noise. 
We note that the same effect could have been achieved by adding small-scale fluctuations to the velocity field. However, since only one of our methods directly uses the velocity field, we opt to be consistent and add small-scale noise to only the density field. We also note that one of the models for generating synthetic \Lya\ skewers (see Model 1 in Table~\ref{tab:models}) does not include lognormal noise. The effect of turning off the extra noise is visible in the power spectrum measurements shown in Fig.~\ref{fig:power_tng}. 

\subsection{Deriving the observed optical depth}
\label{sec:opt}

To convert the fluctuations in the density field into optical depth, we adopt two different, but closely related approaches. The first one of them follows the standard FGPA prescription, while the second introduces a small modification to it. We detail the two methods below.
\begin{itemize}
    \item \textbf{Method I:}
    Two key assumptions go into the FGPA approach~\citep{Gunn:1965ApJ...142.1633G}: adiabatic expansion of the gas and photoionization equilibrium in the IGM. The first one implies that the relationship between density and temperature is well approximated by~\citep{Hui:1997MNRAS.292...27H}
    \begin{equation}
        T(\mathbf{x}) \propto \rho(\mathbf{x})^{\gamma-1},
    \end{equation}
    where $\gamma$ is the slope of the temperature-density relation, while the second dictates the connection between the temperature of the gas and the number of neutral hydrogren atoms:
    \begin{equation}
        n_\mathrm{HI}(\mathbf{x}) \propto \rho(\mathbf{x})^2 T(\mathbf{x})^{-0.7}.
    \end{equation}
    Here, $\rho$ is the baryonic matter density~\citep{Hui:1997ApJ...486..599H}. However, we note that in a collisionless dark-matter simulation, we can only access the total matter field, as defined in Section~\ref{sec:dens}, which we assume traces the baryonic field reasonably well. Combining these two equations and noting that the optical depth, $\tau$, is proportional to the neutral hydrogen column density, $n_{\rm HI}$, we arrive at the final form of the FGPA method~\citep{Bi:1997ApJ...479..523B,Croft:1998ApJ...495...44C}
    \begin{equation} 
        \tau(\textbf{x}) = \tau_0  \ \rho(\textbf{x})^\alpha,
        \label{eq:FGPA}
    \end{equation}    
    where $\tau_0$ is the overall normalization resulting from the temperature-density and photoionization rate assumptions and $\alpha \equiv {2-0.7(\gamma-1)}$. In our analysis, $\tau_0$ and $\gamma$ are free parameters chosen according to the description in Section \ref{sec:tune}. As a helpful reference, we note that the value of the temperature-density slope in TNG can be measured to be $\gamma \approx 1.5$ \citep{2022A&A...664A.198G}.

    The final step in converting the matter field, $\rho(x)$, into the ``observed'' optical depth $\tau (s)$ involves converting the real-space $\tau (x)$ into its redshift-space equivalent, $\tau (s)$. In addition to the redshifting of the \Lya\ absorption features due to cosmic expansion, $\lambda_\mathrm{obs}=\lambda_\alpha(1+z)$, with $\lambda_\alpha$ being the \Lya\   wavelength and $z$ the absorption redshift, there is an additional effect of RSDs caused by the peculiar velocities of neutral hydrogen clouds. We introduce RSDs into our mocks by treating each cell in our three-dimensional grid as a gas blob with a mean velocity along the line of sight as calculated in Section~\ref{sec:dens}. The conversion to redshift-space of each skewer can be expressed as an integral over velocity space of the real-space optical depth multiplied by some kernel, $K$:
    \begin{equation}
        \label{eq:conv}
        \tau({s}) = \int \tau({x}) K\Big({s}-{x}-v_r\big({x}\big) (1+z)/H(z)\Big) \mathrm{d}{x},
    \end{equation}
    where $v_r$ is the peculiar velocity along the line of sight, while $x$ and $s$ are the spatial coordinates in real- and redshift-space, respectively. A typical choice for the kernel in \Lya\   mock generation is the Voigt profile, a Gaussian kernel with a Lorentzian term, or the Doppler profile, just a Gaussian kernel, both of which aim to account for the effects of thermal broadening due to the random thermal velocities of the gas atoms. We implement convolution with the Doppler profile as an option in our \textsc{AbacusSummit} mocks, but find that it has little effect on our observables (e.g., the 1D power spectrum), since the width of the kernel is comparable or smaller than the size of the cells. We show this in Appendix~\ref{sec:voigt}. Thus, to simplify our process, we set $K(x) = \delta^D(x)$, where $\delta^D$ is the Dirac delta function, which amounts to shifting the optical depth of each blob according to its peculiar velocity. In practice, we need to adopt some particle deposition technique due to the discreteness of the cells. A standard choice is to employ a nearest-grid point scheme; however, we opt to use TSC, as it is higher-order than CIC and NGP.
    
    \item \textbf{Method II:}
    Similarly to the first method, here we also assume that the optical depth is related to the density field as $\tau(\textbf{x}) \propto \rho(\textbf{x})^\alpha$. However, the main difference is that in this version, we go directly from the particle positions and their velocities to the final ``observed'' optical depth in redshift space. In particular, we compute a weight for each particle given by $\rho_{\rm dm}(\mathbf{x})^{\alpha-1} \times [1 + \delta_{\rm ln}(\mathbf{x})]^{\alpha}$, where $\rho_{\rm dm}(\mathbf{x})$ is the dark matter density field in real space (see Section~\ref{sec:dens}) and $\delta_{\rm ln}(\mathbf{x})$ is the lognormal noise field in real space introduced in Eq.~\ref{eq:noise}. We then displace the line of sight coordinate component of each particle according to its peculiar velocity as follows:
    \begin{equation}
        s = x + v_r (1+z)/H(z) ,
    \end{equation}
    where $H(z)$ is the Hubble parameter at redshift $z$. Adopting TSC interpolation, we deposit the weighted and displaced particles on the three-dimensional grid to obtain the observed optical depth $\tau (s)$. Thus, this method yields the optical depth directly in redshift-space and as such is less computationally intensive. We note that the reason that this approach leads to the correct form is that the usual particle deposition results in a density field $\rho_{\rm dm} \propto (1+ \delta_{\rm dm})$. Therefore, upon weighting and displacing the particles, we arrive at a field behaving as $\left[\rho_{\rm dm}(\mathbf{x})^{\alpha} \times [1 + \delta_{\rm ln}(\mathbf{x})]^{\alpha}\right](s)$.
\end{itemize}
The main difference between the two methods is that \textbf{Method I} treats the individual grid cells as \Lya\   absorption clouds with a velocity determined by the mean in the cell, whereas in \textbf{Method II}, each particle is approximated as an individual absorber of \Lya\   photons. A downside of the first method is that at low densities, averaging the velocities of sparsely distributed particles results in a suppression of the peculiar velocities of dark matter substructures, which translates as a deficiency in the RSD signal. On the other hand, the second method can potentially lead to a stronger RSD signal than the true \lyaf, as the thermal velocities of individual particles will be larger compared with the gas clouds due to the lack of baryonic pressure in the $N$-body simulation. In an idealized scenario, one could consider identifying substructures via some halo-finding algorithm and deriving the absorption cloud velocities from that, but even this method would not be able to capture correctly the underlying physics, as it would lack important gas and baryonic physics.

\subsection{Obtaining the flux skewers}
\label{sec:flux}

Finally, we need to transform the optical depth, $\tau(s)$, into the transmitted flux fraction, $F(s)$, following:
\begin{equation} 
    \label{eq:tau_to_flux}
   F(s) = {\rm exp}\left[ -\tau(s) \right].
\end{equation}
When computing power spectra of the \lyaf, we typically work with the transmitted flux contrast:
\begin{equation}
    \delta_F(s) = \frac{F}{\langle F \rangle} - 1,    
\end{equation}
which is characterized by having a zero mean. As can be seen in Eq.~\ref{eq:FGPA} and Eq.~\ref{eq:tau_to_flux}, the optical depth is saturated in over-dense regions yielding zero flux and hence no information. On the other hand, more information can be gleaned from low- and intermediate-density regions, where there is some absorption but not enough to cause the signal to be saturated.

\subsection{Parameter tuning}
\label{sec:tune}

\begin{table*}
\begin{center}
\begin{tabular}{c c c c c c c c c c c c c c}
 \hline\hline
Model \# & Method & Fit & $\langle F \rangle$ & $\sqrt{{{\rm Var}}[F]}$ & $\tau_0$ & $\sigma_\epsilon$ & $\gamma$ & $n$ & $k_1$ & $\chi^2_{{\rm 1D}}$ & $\chi^2_{{\rm 3D}}$ & $b_{{\rm Ly \alpha}}$ & $\beta_{{\rm Ly \alpha}}$ \\ [0.5ex]
 \hline
1 & Method I & P3D & 0.801 & 0.168 & 0.387 & 0.000 & 1.650 & --- & --- & 242.276 & 61.138 & $-$0.146 & 0.920 \\ [1ex]
2 & Method I & P1D+P3D & 0.811 & 0.187 & 0.391 & 0.772 & 1.450 & 1.000 & 0.063 & 27.536 & 104.887 & $-$0.129 & 0.949 \\ [1ex]
3 & Method II & P3D & 0.825 & 0.212 & 0.385 & 1.696 & 1.500 & 1.500 & 1.000 & 770.975 & 25.186 & $-$0.130 & 2.022 \\ [1ex]
4 & Method II & P1D+P3D & 0.810 & 0.187 & 0.654 & 2.116 & 1.550 & 0.000 & --- & 131.403 & 100.800 & $-$0.126 & 2.330 \\ [1ex]
 \hline
 \hline
\end{tabular}
\end{center}
\label{tab:models}
\caption{Specifications of the four models used in the creation of the \lyaf\ synthetic catalogs. In particular, we indicate the values of the slow and fast parameters, $\gamma$, $n$, $k_1$, $\tau_0$ and $\sigma_\epsilon$, defined in Section~\ref{sec:noise} and Section~\ref{sec:opt}. Descriptions of the two FGPA-based methods (I and II) can be found in Section~\ref{sec:opt}, while the fitting procedure is detailed in Section~\ref{sec:tune}. The target values of the mean and variance for these mocks are derived from the hydro simulation TNG300-1 and are $\langle F \rangle = 0.8101$  $\smash{({\rm Var}[F])^{1/2} = 0.1878}$. {Model 1 has effectively no small-scale noise added (hence the blanks), while in the case of Model 3, we effectively add ``white'' noise, i.e. with no scale-dependence.} We also share the measurements of the bias and the redshift distortion parameter, $b_{\rm Ly\alpha}$ and $\beta_{\rm Ly\alpha}$, for each model and compare them with the TNG values of -0.1379 and 1.432, respectively. The number of data points fitted in the 1D power spectrum is 27, while that for the 3D power is 20, suggesting that Model 2 provides a good fit for the 1D power, while Model 3 does well with matching the 3D power. We note that some state-of-the-art hydro simulations report higher $\beta_{\rm Ly\alpha}$ values (e.g., Chabanier et al. in prep. find $\beta_{\rm Ly\alpha}=1.8$), which agree better with our Models 3 and 4.}
%{while others report yet lower values \citep[e.g.,][find $\beta \sim 1.3 - 1.5$]{2015JCAP...12..017A}.}, which agree better with our Models 1 and 2.}
\end{table*}

Our model consists of a number of free parameters defined in Section~\ref{sec:noise} and Section~\ref{sec:opt}, namely, $\tau_0$, $\sigma_\epsilon$, $\gamma$, $n$, $k_1$ (see Eq.~\ref{eq:FGPA} and Eq. \ref{eq:noise} for descriptions). To decide on the values of these parameters, we aim to match several key properties of the hydro simulation \lyaf\ skewers: the mean transmitted flux fraction $\langle F \rangle$, the variance of the low-pass-filtered flux with a cutoff at $k_{\rm 1D} = 1 \hompc$, $\langle F^2 \rangle$, the 1D power spectrum, $P_\mathrm{1D}$ (or equivalently, $P(k)$), up to $k < 0.8 \hompc$, and the 3D power spectrum, $P_\mathrm{3D}$ (or equivalently, $P(k, \mu)$), up to $k < 0.8 \hompc$ for selected values of $\mu$. We note that we tune our model by comparing the FGPA-derived skewers on TNG300-3-DM with the full-physics TNG300-1 skewers, as the two simulations share many properties such as initial seed and cosmic variance, which enables a direct comparison. We prefer matching the full shape of the power spectra rather than a compressed statistics such as the bias. The reason for this choice is that due to the limited sizes of the box, low-wavemode quantities are noisy to measure. Only once we are satisfied with the match between TNG300-1 and TNG300-3-DM, do we apply our method to the large boxes of \textsc{AbacusSummit} to obtain the final  products. In addition, we find that the shape of the FGPA-derived power spectra also differs across the different models, which is an additional advantage of matching to the 1D and 3D power spectra. We describe our process in more detail below.

\begin{itemize}
    \item We first measure the 1D and 3D power spectra from TNG300-1 and quantify their error bars. In the case of the 1D power spectrum, the process is straightforward: we Fourier transform the flux contrast $\delta_F$ along each skewer and compute the power spectrum, averaging over all lines-of-sight. {We bin the 1D power spectrum into 400 linear bins ranging from $k \in \{0, \ 12.26\hompc \}$, i.e. spaced by $(2 \pi)/L_{\rm box}$.} To obtain the error bars on the 1D measurement, we apply jackknifes on the available skewers. In the case of the 3D power spectrum, we work with the quantity $P(k, \mu)$ defined in Eq.~\ref{eq:pkmu}. As before, we bin the power spectrum into 20 $k$ bins ranging from $k \in \{(2 \pi)/L_{\rm box}, \ 15\hompc \}$ and 16 $\mu$ bins ranging from 0 to 1. We assume that the error bars on this measurement are well approximated by the Gaussian error:
    \begin{equation}
        \Delta P(k, \mu) = \sqrt{\frac{2}{N_k}} P(k, \mu),    
    \end{equation}
    where we calculate the number of k modes in each $k$ and $\mu$ bin as $N_k = k^2 dk d\mu / (2 \pi/L_{\rm box})^3$ with $L_{\rm box}$ being the box size. 
%We have checked explicitly that the Gaussian approximation holds by splitting one of the large \textsc{AbacusSummit} boxes into 64 subboxes and performing jackknife sampling. 
    
    We next split the tuning process into a slow and a fast step, with the fast step varying $\tau_0$ and $\sigma_\epsilon$ to match the mean and variance of the flux, and slow step varying $\gamma$, $n$, and $k_1$ to additionally match the 1D and 3D power spectra. 
    \item \textbf{Fast parameters:} For a given choice of slow parameters, $\gamma$, $n$, and $k_1$, and a Method (I or II as defined in Section~\ref{sec:opt}), we vary the parameters $\tau_0$ and $\sigma_\epsilon$, so as to minimize the function:
    \begin{eqnarray}
        \chi^2_{\rm mean, std} = \big[\langle F_{\rm TNG \ Hydro} \rangle - \langle F_{\rm TNG \ FGPA} \rangle \big]^2 + \\ \big[\sqrt{{\rm Var}[F_{\rm TNG \ Hydro}]} - \sqrt{{\rm Var}[F_{\rm TNG \ FGPA}]} \big]^2 ,
        \nonumber
    \end{eqnarray}
    where the mean flux $\langle F_{\rm TNG \ Hydro} \rangle = 0.8101$ is taken from the empirical relation, i.e. Eq.~\ref{eq:mean}, while the variance, $\smash{({\rm Var}[F_{\rm TNG \ Hydro}])^{1/2} = 0.1878}$, is computed from the low-pass filtered flux skewers described above. We note that the fast parameters are optimized separately from the slow ones, which is the reason that we do not worry about normalizing the $\chi^2$. Furthermore, we implicitly assume that the error on the mean and standard deviation measurements is comparable. {We test this assumption for our default resolution and find that the two differ only by a small $\mathcal{O}(1)$ factor.}
    \item \textbf{Slow parameters:}
    To decide on the values of the slow parameters, we generate a three-dimensional uniform grid with possible values they can take: we allow $\gamma$ to vary between 1.4 and 1.7 and test 6 values in that range (typically, one assumes that $\alpha \approx 1.6$, which corresponds to $\gamma \approx 1.56$); $n$ is allowed to vary between -3 and 3, and we test 200 values in that range; finally, $k_1$ varies between 0.001 to 1, and we test 200 values in that range. We do not sample $\gamma$ as  densely as the other two parameters, as we find that our observables are weakly affected by this choice.

    For each of the two methods (see Section~\ref{sec:opt}) and each point in the three-dimensional grid, we first fit for the mean and variance of the flux so as to calibrate $\tau_0$ and $\sigma_\epsilon$ and then record the mean and the flux alongside the contribution to the $\chi^2$ of the 1D and 3D power spectra, computed as follows:
    \begin{eqnarray}
        \chi^2_{\rm 1D} = {\sum_k} \left[ \frac{(P_{\rm 1D, TNG \ Hydro}(k) - P_{\rm 1D, TNG \ FGPA} (k))}{\Delta P_{\rm 1D, TNG \ Hydro} (k)} \right]^2 \\
        \chi^2_{\rm 3D} = {\sum_{k, \{\mu\}}}\left[ \frac{(P_{\rm TNG \ Hydro}(k, \mu) - P_{\rm TNG \ FGPA}(k, \mu))}{\Delta P_{\rm TNG \ Hydro} (k, \mu)} \right]^2
        \nonumber
    \end{eqnarray}
    for four selected values of $\mu$, namely $\{ 0.03,\ 0.33,\ 0.66,\ 0.97\}$ with a bin width of $\Delta \mu \approx 0.06$. We have checked that the parameter selection is negligibly affected by whether we use only a handful of $\mu$ values or the full $P(k, \mu)$ vector.
    
    Note: The reason we classify $n$ and $k_1$ as slow parameters is that the step of generating the Gaussian noise skewers is relatively slow. Similarly, changing $\gamma$ in \textbf{Method II} requires a rerunning of the TSC particle deposition step, which is computationally expensive. 
    
    \item In the final step of this process, we select the values of the three slow parameters, which will be used in the \textsc{AbacusSummit} \lyaf\ mocks. To do so, we combine the $\chi^2$ values from the 1D and 3D power spectra. In particular, for each of the two methods (see Section~\ref{sec:opt}), we choose two sets of slow parameters: the first set corresponds to the best-fit parameters we obtain when minimizing $\smash{\chi^2_{\rm Model \ 1} = \chi^2_{\rm 3D}}$, while the second set comes from minimizing $\smash{\chi^2_{\rm Model \ 2} = \chi^2_{\rm 3D} + \chi^2_{\rm 1D}}$. We quote the best-fit values for all four models (two per method) in Table~\ref{tab:models}. We note that we do not include the mean and variance $\chi^2$ contributions, {as those are already calibrated individually for each model. Additionally, since the errors on the mean and the variance are much smaller, they would dominate the selection, and our final objective is to match the power spectra.}
    
    {We note that while the values of the fast parameters $\tau_0$ and $\sigma_\epsilon$ in the large boxes are quite similar to the TNG-DM boxes, we opt to recalibrate them to make sure we match the mean and variance of the flux as best as we can.}
\end{itemize}

{To summarize the tuning process, we start by generating a three-dimensional regular grid, for which each point corresponds to a set of predetermined values for the three slow parameters. For each set of three slow parameters, we minimize the absolute difference with TNG300-1 of the mean and the variance of the flux, adopting the Nelder-Mead scheme, to find the values of the fast parameters and record the resulting 1D and 3D power spectrum difference (in terms of $\chi^2$). For each of the 4 models considered in this work, we then simply report the set of slow and fast parameters that correspond to the smallest $\chi^2$ across all grid points. In the future, we plan to adopt a more flexible iterative process rather than the preset three-dimensional grid. However, that would require that we substantially speed up the power spectrum computation, for example, by adopting analytical approximations \citep[e.g.,][]{2020JCAP...03..068F}. We defer these ideas for later work, where we explore a more complex model and utilize larger boxes for calibration.}
%(in that case, the efficiency of the tuning process will be essential).

\begin{figure}
    \centering
    \includegraphics[width=0.48\textwidth]{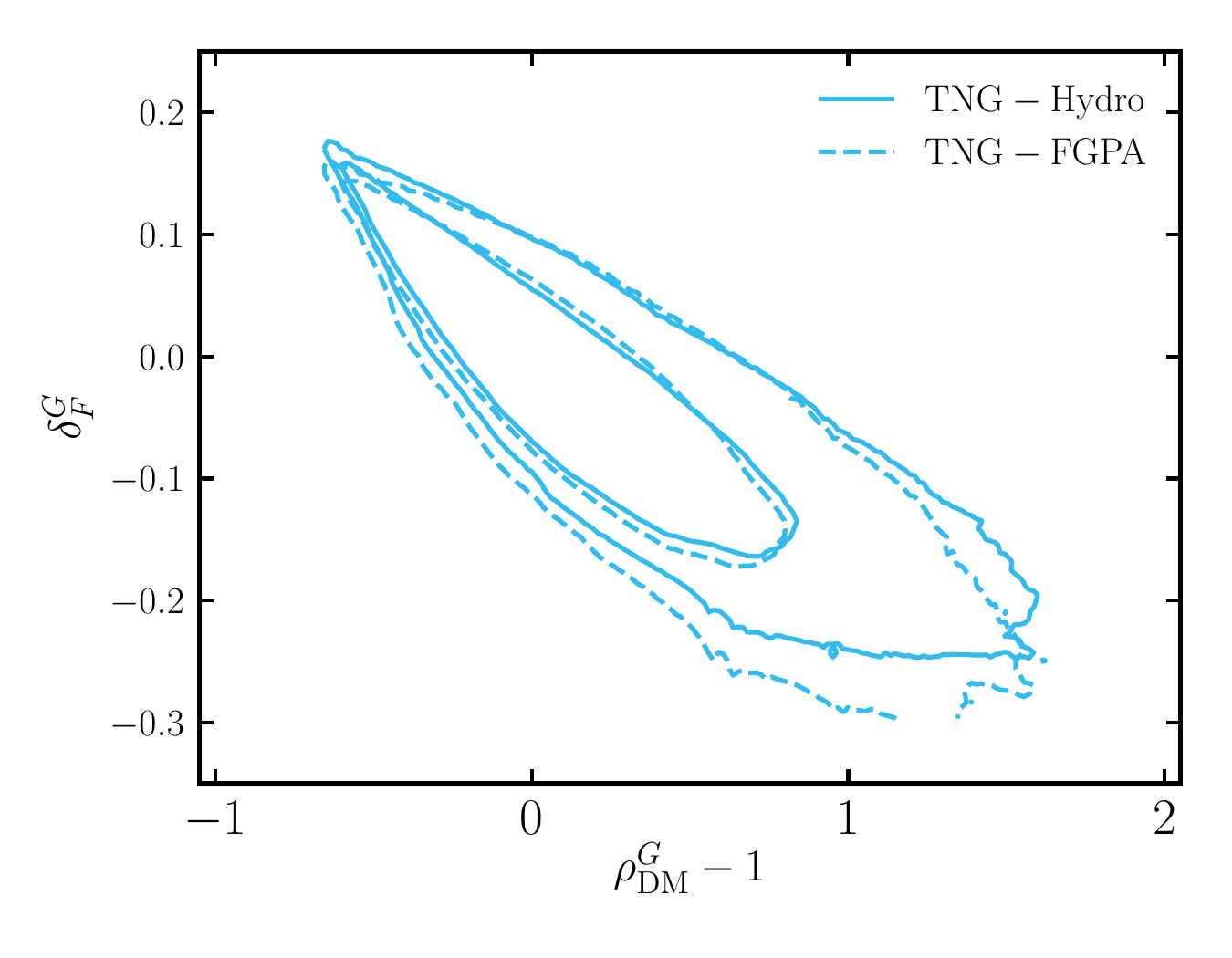}
    \caption{Two-dimensional PDF contours comparing the dark-matter-field-Ly$\alpha$-flux relation for the TNG300-1 hydro simulation (solid) and one of our FGPA-based synthetic catalogs (dashed) applied to the low-resolution dark-matter-only counterpart TNG300-3-DM (Model 3; see Table~\ref{tab:models}). The levels shown correspond to 2\% and 68\%. The voxel resolution of the maps is $0.33\mpcoh$, and both maps are smoothed with a Gaussian kernel of size $\sigma_G = 3\mpcoh$ for visualization purposes. The similarity between the two curves confirms that the gas physics has small effect on Ly$\alpha$ observables on megaparsec scales.} 
    \label{fig:flux_dens}
\end{figure}

In Fig.~\ref{fig:flux_dens}, we show the two-dimensional PDF contours comparing the dark-matter-field-Ly$\alpha$-flux relation for the TNG300-1 hydro simulation and one of our FGPA-based synthetic catalogs applied to the low-resolution dark-matter-only counterpart TNG300-3-DM (Model 3; see Table~\ref{tab:models}). The levels shown correspond to 2\% and 68\%. The voxel resolution of the maps is $0.33\mpcoh$ and both are smoothed with a Gaussian kernel of size $\sigma_G = 3\mpcoh$. Note that the smoothing is applied for visualization purposes and is not used for any other figure in this paper. The similarity between the two curves confirms that the gas physics has small effect on Ly$\alpha$ observables on megaparsec scales.

\begin{figure*}
    \centering
    \includegraphics[width=0.98\textwidth]{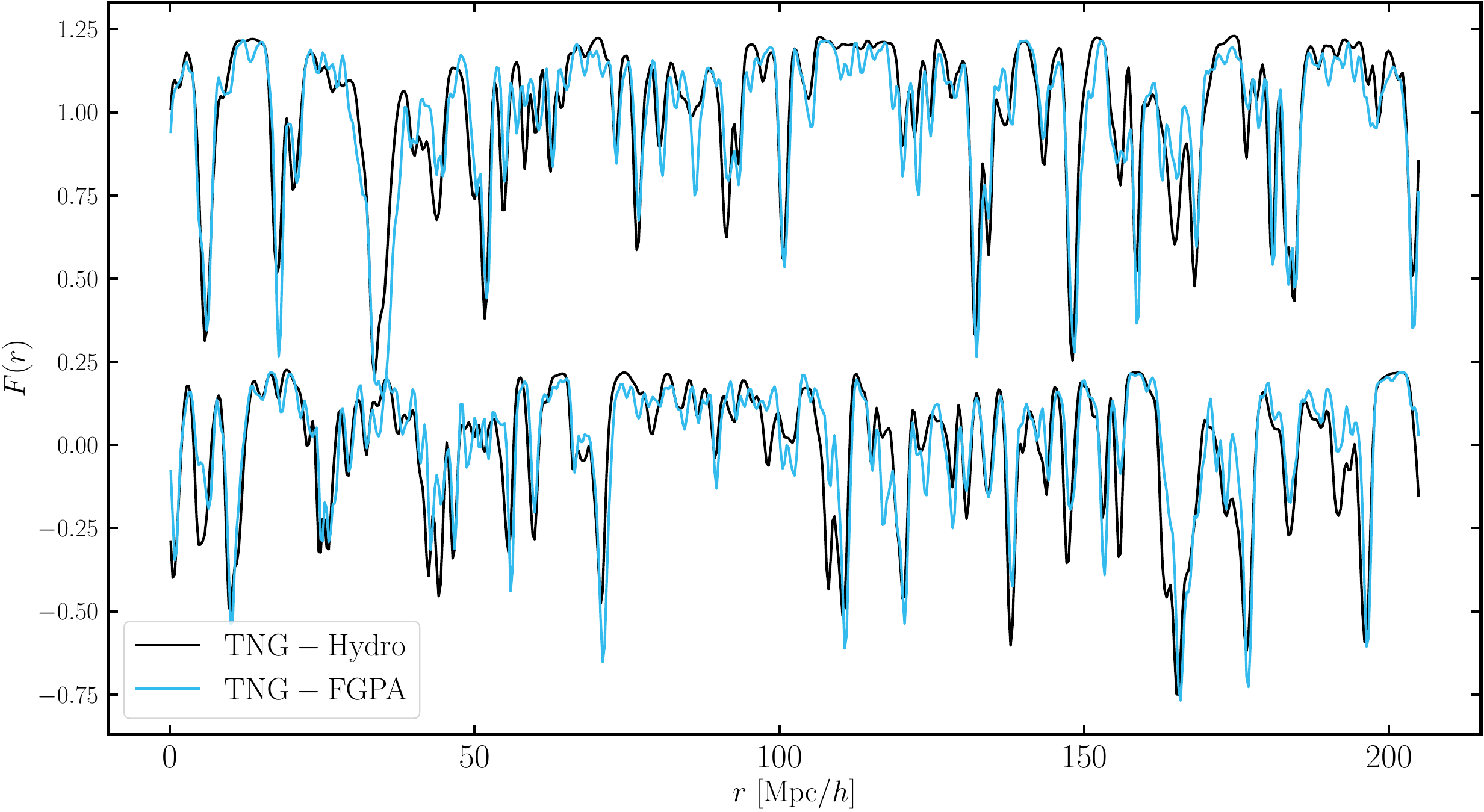}
    \caption{Skewers of the ``true'' \lyaf\ flux (solid) and the FGPA prediction (dashed) for our Model 3 (see Table~\ref{tab:models}). It is reassuring to see that the simplified FGPA model does an adequate job of matching the majority of the features present in the full hydrodynamical spectra. For easier visualization, we are plotting $\delta F(r) + 1 \equiv F(r)/\bar F(r)$ and $\delta F$ for the skewer on top and bottom, respectively.} 
    \label{fig:skewers}
\end{figure*}

In Fig.~\ref{fig:skewers}, we show a couple of skewers passing through the entire TNG300 box for the ``true'' \Lya\   spectra extracted from TNG300-1 and the synthetic ones obtained using our Model 3 (see Section~\ref{sec:mocks} and Table~\ref{tab:models}) on the low-resolution counterpart TNG300-3-DM. For visualization purposes, we plot $\delta F(r) + 1 \equiv F(r)/\bar F(r)$ for the skewer on top and $\delta F$ for the skewer on the bottom. 
Reassuringly, the simplified FGPA model does an adequate job of matching the majority of the features present in the full hydrodynamical spectra.
%, tracing well the large- and intermediate-scale fluctuations, which contain the majority of the cosmological information in the \lyaf. 
Visible in the comparison of the two is that the true skewers appear smoother than the FGPA ones due to the extra noise added to the latter. 
%On the other hand, Model 3 adds a small amount of uncorrelated small-scale power beyond $k > 1 \hompc$, which can be seen by eye as tiny jaggedy features. 
As we will see in Fig.~\ref{fig:power_tng}, the Model 3 FGPA 1D power spectrum overshoots the true 1D power spectrum partly due to the addition of small-scale power. Reassuringly, we have inspected (not shown) the skewers for Model 1, which has no small-scale power added, and found the opposite: the FGPA skewers lack small-scale features compared with the true skewers (and their 1D power spectrum, shown in Fig.~\ref{fig:power_tng}, is lower, as expected).

\subsubsection{Comparing the power spectrum of TNG300-1 and TNG-300-3-DM} 
\label{sec:tng_tngdm}

\begin{figure*}
    \centering
    \includegraphics[width=0.98\textwidth]{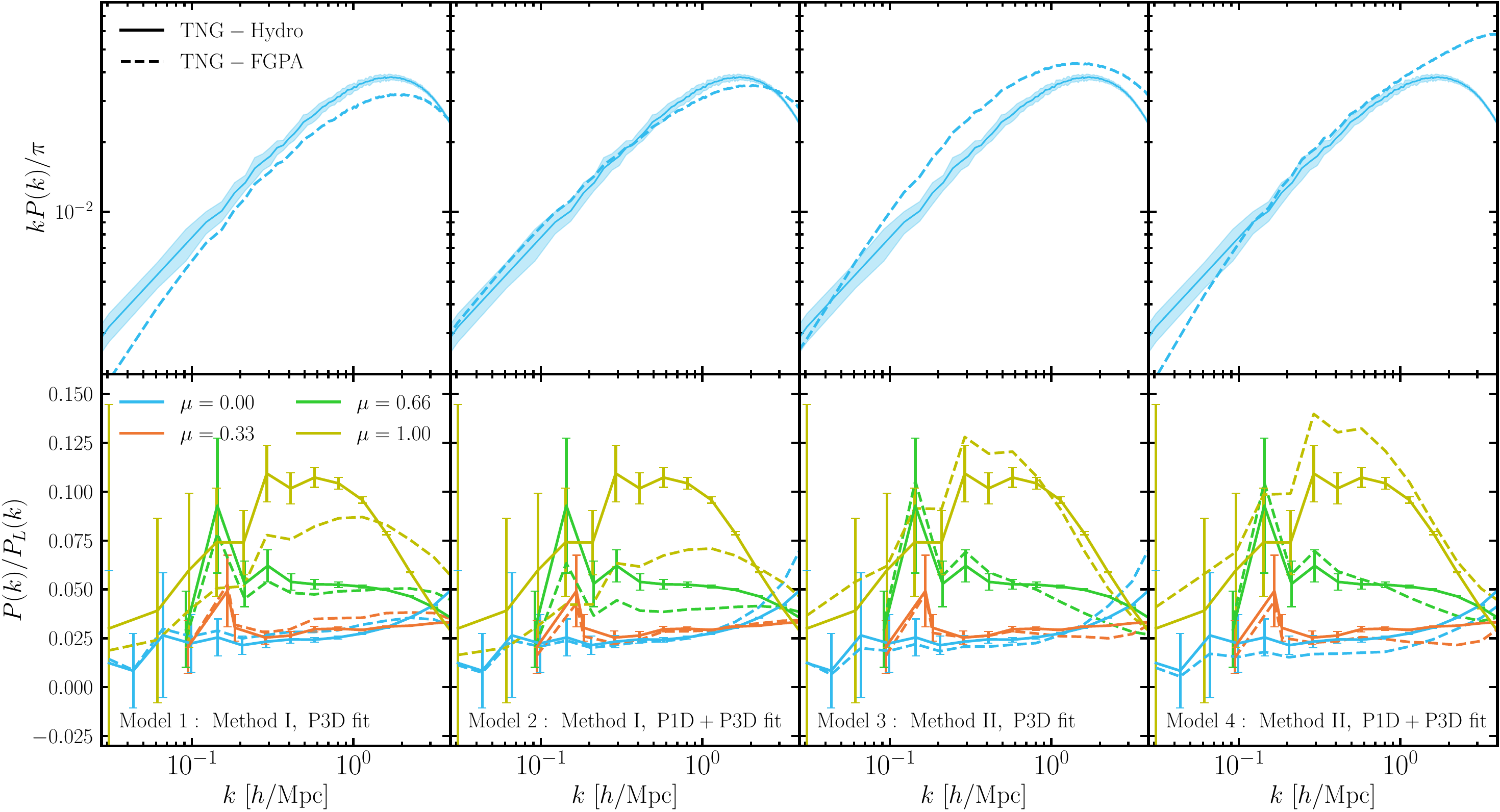}
    \caption{1D and 3D power spectrum of the \lyaf\ generated for the four models presented in this work (see Table~\ref{tab:models} and Section~\ref{sec:mocks}) when applied to the $N$-body simulation TNG300-3-DM (dashed curves) and the ``true'' measurements from the hydro simulation (solid curves). In the top panels, we show the 1D power spectrum, $P_{\rm 1D}(k)$, with the shaded regions indicating jackknife errors, whereas in the bottom ones, we show the ratio between the linear power spectrum and $P(k, \mu)$ for four different values of $\mu = \{0, 0.33, 0.66, 1\}$, with the error bars coming from the Gaussian prediction. Note that the sample variance is the same in both boxes, which facilitates the comparison, and that on small scales, the measurement is affected by the effects of interlacing and aliasing due to the sparseness of the lines-of-sight \citep{2007PhRvD..76f3009M} and the cell size. Overall, the four models exhibit a good agreement with the ``truth'' and provide a wide variety of synthetic catalogs for the users of these mocks.} 
    \label{fig:power_tng}
\end{figure*}

We produce \textsc{AbacusSummit} \lyaf\ mocks for four different models: two FGPA-based methods (see Section~\ref{sec:opt}) calibrated to match the 3D power spectrum individually and the 1D and 3D power spectra jointly of the TNG300-1 skewers. In Fig.~\ref{fig:power_tng}, we illustrate the level of agreement between the hydro run TNG300-1 and the FGPA mocks run on the dark-matter-only simulation TNG300-3-DM. We note that the sample variance is the same in both boxes, which facilitates the comparison of the models to the ``truth.'' In addition, on small scales, the measurement is affected by an interlacing effect due to the size of the cells and an aliasing effect due to the sparseness of the lines-of-sight, which scales as $P_{\rm 1D}/n_{\rm 2D}$ \citep{2007PhRvD..76f3009M}, where $n_{\rm 2D}$ is the two dimensional density of skewers. {Subtracting the aliasing effect theoretically is not trivial, as the skewers in our mocks are placed in a regular grid rather than randomly, as would be the case in observations, and thus the formula in \citet{2007PhRvD..76f3009M} does not hold.}
%is unknown. To account for this effect properly, one needs to fit for the amplitude given the measured power spectrum, similarly to the shotnoise fits performed when studying galaxy clustering. 
In this work, we opt not to do this, as the affected $k$-modes are beyond our scales of interest. Similarly, interlacing affects our measurements beyond $k \sim 4-5 \hompc$ and can be numerically corrected by offsetting the grid by half a cell size and recomputing the Ly$\alpha$ observables. As this is prohibitively expensive in the case of the AbacusSummit boxes, which are generated using a lightweight, single-node script, we choose not to apply it to the TNG case either, as we try to make the comparison as consistent as possible (for example, by choosing similar resolution and grid size).

 As expected, Models 2 and 4, which are aiming to fit both the 1D and the 3D power spectrum, exhibit closest agreement to the ``true'' (TNG300-1) 1D power spectrum for $k < 1 \hompc$. {Model 4 over-predicts the 1D power on the smallest scales, not included in the fits, probably because of the large amount of extra power added (large value of $\sigma_\epsilon$).} 
% We attribute this to the fact that the dark-matter simulation lacks the effect of baryonic pressure, which would naturally have suppressed the signal on small scales.
 %Model 2 (produced via Method I, see Section~\ref{sec:opt}) suffers less from this effect, since the velocities (and thus effective pressure) of the gas blobs are already diminished due to the TSC interpolation of the peculiar velocity field (see Section \ref{sec:dens}). 
 %{As a result of the suppression of the velocity field fluctuations when adopting TSC painting in Method I, both Models 1 and 2 have lower 3D power on intermediate scales. The LyaCoLoRe mocks \citep{2020JCAP...03..068F} address this issue by artificially boosting the velocities by 30\%. 
 %Models 3 and 4 do not suffer from this issue, as the RSDs are applied directly to the particles rather than the voxels. 
 
 {In terms of the 3D power, as expected, Models 1 and 3 show better agreement with TNG300-1, respectively, than Models 2 and 4. Models 1 and 2, moreover, have noticeably smaller redshift-space distortions (lower value of the parameter $\beta$). This was also the case in the FGPA-lognormal mocks presented in \citet{2020JCAP...03..068F}, where the authors addressed this issue by artificially boosting the velocities by 30\%.}
 %, with Model 3 predicting higher 1D power spectrum.} 
 %Curiously, its shape is reminiscent of the observational measurement from eBOSS shown in Fig.~\ref{fig:p1d_eboss}, suggesting that this model might yield more similar measurements to the observations. 
 Overall, the four models exhibit a reasonable agreement with the hydro ``truth,'' providing a wide selection of synthetic catalogs for the users of these mocks to have at their disposal.

\section{Validation of the mocks}
\label{sec:stats}

In this Section, we study observable summary statistics of our \lyaf\ mocks relevant for current and future surveys, namely, the 1D and 3D power spectrum, and the correlation function, in order to validate our mocks. In particular, we first introduce the available large-volume synthetic products on \textsc{AbacusSummit}. We then show measurements of these statistics from our mock skewers and discuss their shortfalls and successes in recovering the ``true'' statistics coming from the hydro simulation, IllustrisTNG. We then compute the real-space clustering of our \textsc{AbacusSummit} \lyaf\ skewers and study the effect of non-linear broadening on the BAO peak. We also compare our measurements against observations from eBOSS \citep{2020ApJ...901..153D}.
%\citep{2017A&A...608A.130D,2020ApJ...901..153D}.

\subsection{Available \textsc{AbacusSummit} products}
\label{sec:prod}

All \textsc{AbacusSummit} \lyaf\ mocks are generated on a single node of the National Energy Research Scientific Computing (NERSC) Centre's \texttt{cori} machine using specially developed \texttt{python} scipts with no external dependencies apart from \texttt{scipy}, \texttt{numba}, and the specialized package for reading \textsc{AbacusSummit} products, \texttt{abacusutils}\footnote{The package and instructions for installing it can be found here: \url{https://github.com/abacusorg/abacusutils}.}. The maximum RAM consumption is capped at 70 GB for any of the scripts and the total size of all products (6 simulations, 4 models, 2 lines-of-sight) after applying ASDF `blsc' compression is 50 TB\footnote{Available via the package \texttt{abacusutils}.}. As discussed in Section~\ref{sec:tune}, we generate mocks for four separate models (adopting \textbf{Method I} and \textbf{II} to fit the 1D power spectrum and the 1D+3D power spectrum, subsequently). Our products are available for each of the six fiducial cosmology simulations \texttt{AbacusSummit\_base\_c000\_ph000-005} ($2\gpcoh$, 6912$^3$) at $z = 2.5$ and each of the four models, and can be downloaded via Globus (see Data Availability). Each \lyaf\ mock has a resolution of 0.29 $\mpcoh$ per cell, corresponding to a total of 6912$^3$ grid cells. Below, we list the specifications of our \lyaf\ and QSO catalogs for each simulation:
\begin{itemize}
    \item Two full sets of redshift-space optical depth skewers (6912$^2$, with 6912 line-of-sight pixels), one placing the observer along the $y$ axis and one along the $z$ axis\footnote{We do not generate maps with the line-of-sight direction being along the $x$ axis, as the \textsc{AbacusSummit} particle outputs are split into slabs along the $x$-axis that we analyze independently for the sake of efficiency.}. These skewers can be easily converted into flux transmission skewers according to Eq.~\ref{eq:tau_to_flux}. Each map takes up 1 TB of disk space and is split into 144 pieces each containing 48$\times$6912 lines-of-sight.
    \item Two sets of complex $\tilde \delta_F (\mathbf{k})$ maps (as before, provided for line-of-sight along $y$ and $z$ directions) generated by Fourier transforming the flux contrast field $\smash{\delta_F (\mathbf{x})}$ (6912$^3$ cells) and then low-pass filtering the result, i.e. removing the small-scale modes, $\smash{k_{\rm max,los} > 4 \hompc}$ and $k_{\rm max,perp} > 2 \hompc$ along and perpendicular to the line-of-sight, respectively\footnote{In order to perform the Fourier transform of a 6912$^3$ map on a single node, we consecutively load each slab in $x$, apply Fourier transformation in $y$ and $z$ and then low-pass filter the resulting array, until we are finished with all slabs and can apply one final low-pass filter along $x$.}. We filter out small scales, which we know are dominated by baryonic effects missing in our simulations, to save disk space (each of the Fourier maps is 13 GB). We note that DESI will measure the \lyaf\ power spectrum down to 2-3 $\hompc$, so these complex maps provide sufficient small-scale information for modeling the DESI measurements.
    \item QSO catalogs containing the positions, velocities and host halo masses of each quasar with RSD effects applied along the $y$ and $z$ axis. The sample is generated via \textsc{AbacusHOD} as described in Section~\ref{sec:quasar} with a number density of $1.75\times 10^{-4} \ [\mpcoh]^{-3}$ (i.e., 1.4 million quasars per box) and a bias and redshift distortion parameter of $b^{\rm QSO} \approx 3.3$ (i.e., $\smash{\beta_{\rm QSO} = f/b_q \approx 0.294}$), chosen to be close to the eBOSS measurement \citep{2020ApJ...901..153D}.% of $b^{\rm QSO} \approx 3.7$ (i.e., $\beta_{\rm QSO} = f/b_q \approx 0.26$) at $z_{\rm eff} = 2.334$ and the expected values for the DESI quasars \citep{2016arXiv161100036D}.
    \item Similarly to the complex maps we generate for the \lyaf\ quantities, $\tilde \delta_F (\mathbf{k})$, we also generate complex maps of the quasar overdensity field, $\smash{\tilde \delta^{\rm QSO}_g(\mathbf{k})}$, calculated by Fourier transforming the quasar overdensity field, $\smash{\delta^{\rm QSO}_g(\mathbf{x})}$, obtained through the TSC interpolation of the redshift-distorted quasar positions on the three-dimensional grid (6912$^3$ cells), and applying a low-pass filter of $k_{\rm max,los} < 4 \hompc$ and $k_{\rm max,perp} < 2 \hompc$. 
\end{itemize}

\subsection{Power spectrum}
\label{sec:power}

As described in Section~\ref{sec:tune}, when deciding on the values of the free parameters introduced in our model, we try to maximize the similarity between the power spectrum measurements from the ``true'' \lyaf\ skewers extracted from TNG300-1 and TNG300-3-DM equipped with our two FGPA-based methods (see Section~\ref{sec:opt} for descriptions). As a reminder, the 1D power spectrum of the \lyaf\ is measured by Fourier transforming each skewer along the line-of-sight and averaging over all lines-of-sight to arrive at the final quantity. Thus, each skewer is treated independently and this statistic does not take into account any cross-correlation between different lines-of-sight. On the other hand, the second statistic, $P(k, \mu)$ (see Eq.~\ref{eq:pkmu}), incorporates the correlation between skewers: $P(k, \mu = 0)$ measures the power in the transverse direction, whereas $P(k, \mu = 1)$ measures it in the direction parallel to the line-of-sight.

The end goal of this project is to generate \lyaf\ mocks in volumes sufficiently large to aid the analysis of large-scale surveys targeting quasars such as DESI. For this reason, it is of utmost importance that we can scale up our algorithm and run it successfully on the $2\gpcoh$ \textsc{AbacusSummit} boxes. We note that as the cell grid of the data increases substantially between TNG300-1-DM and \textsc{AbacusSummit}, it is necessary to refactor and rewrite our scripts altering the straightforward implementation described in Section~\ref{sec:mocks}. Therefore, verifying that our method yields results comparable to TNG300-3-DM is an essential step before our mocks are declared satisfactory. An additional complexity is that the resolution of TNG300-3 (mean particle distance of $0.33\mpcoh$) and the \textsc{AbacusSummit} \texttt{base} boxes differs slightly (mean particle distance of $0.29\mpcoh$). Ideally, one would want to recalibrate the slow parameters (see Section~\ref{sec:tune} for a definition of ``slow'' vs. ``fast'') for each distinct simulation, but that would constitute a substantial computational burden. Here, we demonstrate that the behavior of the \textsc{AbacusSummit} mocks is sufficiently similar given our targeted precision, so we defer a more complex treatment to future work.

\begin{figure*}
    \centering
    \includegraphics[width=0.98\textwidth]{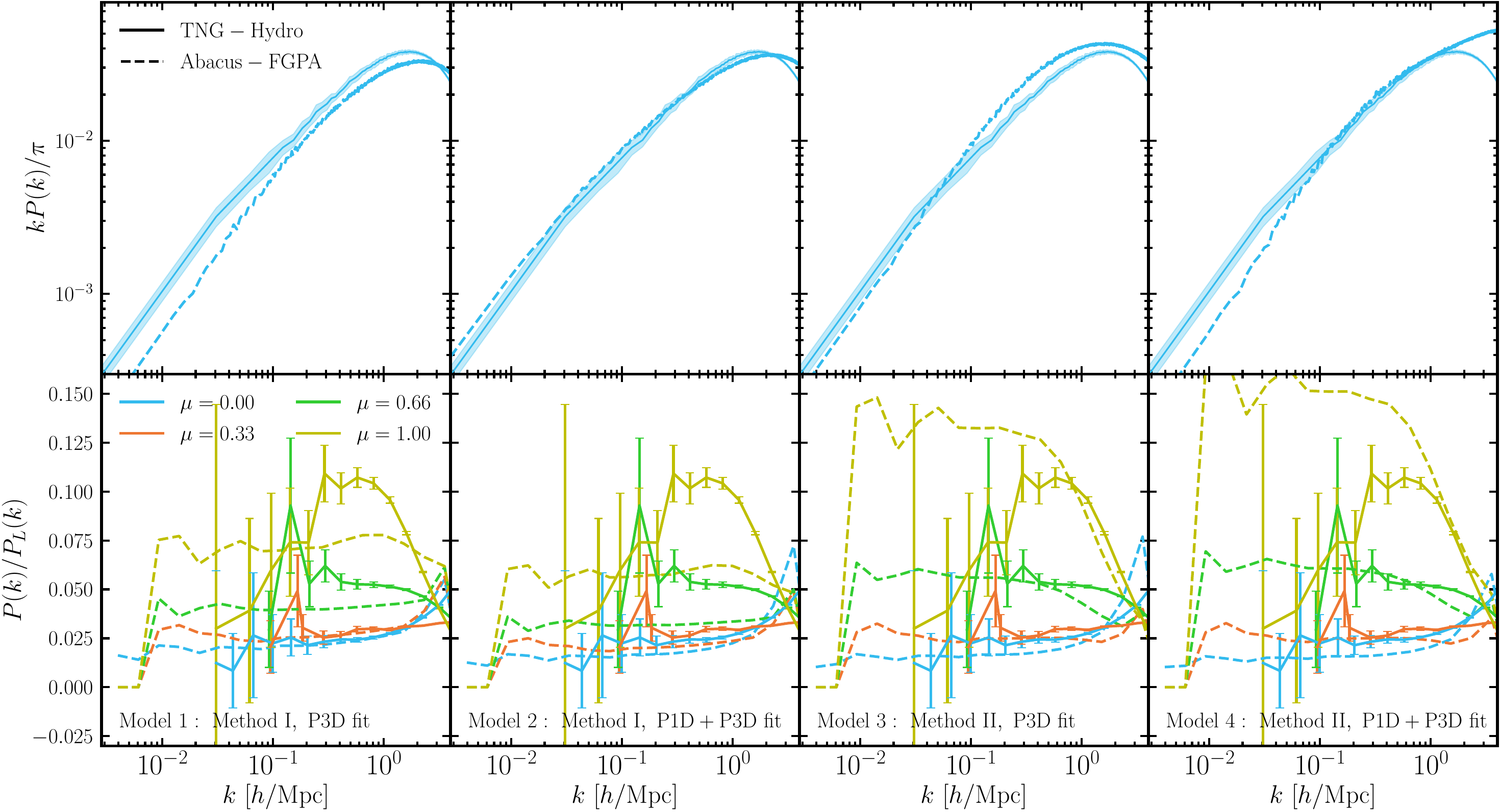}
    \caption{Similarly to Fig.~\ref{fig:power_tng}, 1D and 3D power spectrum of the \lyaf\ skewers extracted from our \textsc{AbacusSummit} mocks using four different models (see Table~\ref{tab:models} and Section~\ref{sec:mocks}). In this case, the $N$-body simulation (\texttt{ph000}) is much larger than the hydro one ($2\gpcoh$ vs. $205\mpcoh$), which allows us to extend the power spectrum measurements by about an order of magnitude to larger scales. We cut the smallest scales shown to $k < 4\hompc$, as for this measurement, we use the complex $\delta_F$ maps, which are available up to $k_{\rm max, los} = 4 \hompc$. Reassuringly, the agreement with TNG300-1 is comparable to our findings in Fig.~\ref{fig:power_tng} and provides us with confidence as to the validity of our \lyaf\ mocks.} 
    \label{fig:power_abacus}
\end{figure*}

In Fig.~\ref{fig:power_abacus}, we study the 1D and 3D power spectrum of the ``true'' \lyaf\ skewers from TNG300-1 and the skewers obtained for each of our four models from Section~\ref{sec:tune} applied to \textsc{AbacusSummit}. We find that the agreement of our mocks with TNG300-1 is similar to the agreement between TNG300-1 and TNG300-3-DM (see Fig.~\ref{fig:power_tng}). We cut the smallest scales shown to $k < 4 \hompc$, as for these measurements, we employ the complex $\delta_F$ maps, which are available up to $k_{\rm max, los} = 4 \hompc$ and easier to handle than the real-space $\tau$ skewers. It is reassuring that the agreement with TNG300-1 is comparable to our findings in Fig.~\ref{fig:power_tng}, suggesting that the implementation of the mocks in the larger \textsc{AbacusSummit} boxes has been successful. Remaining differences in the intermediate regime, shared by both TNG300-3-DM and \textsc{AbacusSummit} can be attributed to differences in the resolution and the cosmological parameters.

{We perform an additional test of dividing the power spectrum by the linear theory prediction with matching best-fit bias and $\beta$. We find that the mock power spectra agree within 10\% with the linear theory result up to $k \lesssim 0.4 \hompc$, after which they begin to diverge noticeably from linear theory. The agreement within the Method I Models (i.e., 1 and 2) and within the Method II Models (i.e., 3 and 4) is excellent until $k \lesssim 2 \hompc$, but the two methods show larger deviations between each other beyond $k \sim 0.7 \hompc$, especially for high $\mu$ values.}

%{This point is also demonstrated using the correlation function in Fig.~\ref{fig:multipole_theory} and further discussed in Section~\ref{sec:comp_lin}.}
%AFR: I'm not sure about the connection between this discussion and Fig.~\ref{fig:multipole_theory}. Maybe you mean \ref{fig:multipole_eboss}?}  
%BH: Yes, sorry, meant Fig. 7; let's just remove.
%{The value of our mocks is that they will offer a path forward for theorists to robustly test full-shape models on intermediate and large scales, which should be largely insensitive to the particular details of the \lya\ painting technique.}
%AFR: Can we really say this last sentence? We have seen that the details of the \lya\ painting technique (Method I or II) make a big difference on mildly non-linear regime. They don't agree whether the power should be larger or smaller than Kaiser.
%BH: Agreed

%This result indicates that the particular details of our \lya\ painting technique do not matter at the scales of interest for this study ($k \sim 2 \hompc$) and that provided one marginalizes over bias and $\beta$, the mocks provide a realistic prediction of the behavior of \lyaf\ skewers at $k \lesssim 1 \hompc$.
%, which is reassuring, as it suggests that our mocks are robust at these scales.

\subsection{Correlation function}
\label{sec:corr}

Modern surveys will be capable of measuring the spectra of millions of distant objects and make subpercent measurements of the flux decrement correlation function over a wide range of scales and redshifts. This provides a handle of crucial cosmological measurements, such as the angular and redshift scale of the BAO, cosmic expansion, and the effect of neutrinos on the power spectrum. When measuring the small-scale clustering of galaxies, one can directly relate the redshift distortion parameter $\beta$ to the growth of structure; however, in the case of the \lyaf, $\beta$ depends on a second bias factor that must be determined independently, which
%is related to the response of the mean value of F to a large-scale peculiar velocity gradient,
{comes from a more general linear theory calculation of RSDs in which the distorted field, in this case $\tau$, undergoes a non-linear transformation, in this case $F = \exp(-\tau)$ \citep{McDonald2000}.} One viable way of doing so is by jointly analyzing the two-point correlation function of Ly$\alpha$-Ly$\alpha$, Ly$\alpha$-QSO, and QSO-QSO in a ``3$\times$2-pt'' fashion \citep{2021MNRAS.506.5439C}. However, to do so reliably, we need to extensively test our analysis tools on realistic mocks. Hence, this is one of the main objectives of our data products. In addition, it is well known that non-linear evolution causes a broadening of the BAO peak in the correlation function of galaxies. Therefore, it is interesting to ask whether the BAO peak in the flux correlation function is similarly broadened. In this work, we explore the auto- and cross-correlation function of Ly$\alpha$ and QSO and demonstrate the non-linear broadening of the Ly$\alpha$-measured BAO peak for the first time in simulations. This is crucial to incorporate in and test through our theoretical models, as we expect that real Ly$\alpha$ observations will also be affected. %\citep{2012JCAP...11..059F}.

{We summarize the flow of the section here to make it easier for the reader to follow. In Section~\ref{sec:measure_corr}, we sketch out the calculation connecting the theoretical power spectrum $P(k, \mu)$ to the correlation function multipoles, $\xi_\ell$. The utility of this calculation is two-fold: to convert our simulated power spectrum into the simulated correlation function via the Hankel transform, we need a smooth function on very large scales, which we supply via linear theory by fitting the bias and $\beta$ parameters to the simulated $P(k, \mu)$. On the other hand, we want to compare the simulated correlation function near the BAO scale with some theoretical model, so use these equations to calculate the linear theory prediction and also two models of the BAO peak broadening, defined in Section~\ref{sec:comp_lin}.} %Finally, we compare the simulated correlation function with the \texttt{Vega} model used in the eBOSS analysis in Section~\ref{}.}

\subsubsection{Measuring the correlation function from the power spectrum}
\label{sec:measure_corr}

To obtain the two-point correlation function measurement from our mocks, we start by calculating the power spectrum, $P(k, \mu)$, as before (see Eq.~\ref{eq:pkmu}). We adopt maximum $k_{\rm max} = 1.6 \hompc$ to speed up the calculations and because for this part of the validation, we are mostly interested in the BAO scales. We also calculate the multipoles of the redshift-distorted power spectrum, $P(k, \mu)$, via:
\begin{equation}
    P_{\ell}(k) = \frac{2\ell + 1}{2}\, \int_{-1}^{+1} P(k,\mu)\, L_{\ell}(\mu)\,d\mu \; ,
\end{equation}
where $L_{\ell}$ is the Legendre polynomial and $P_{\ell}(k)$ are the multipoles of the redshift-distorted power spectrum $P(k,\mu)$. 

Approximating the error on the measurement as Gaussian, we fit the $b$ and $\beta$ parameters to linear theory with the Kaiser approximation \cite{1987MNRAS.227....1K}:
\begin{equation}
\label{eq:lin}
    \tilde P(k, \mu) = b^2 (1+\beta \mu^2)^2 \ \tilde P(k),
\end{equation}
where the $\tilde P$ signifies this is a theory prediction. Similarly, we fit the cross-power spectrum between Ly$\alpha$ and quasars via
\begin{equation}
    \tilde P_{q}(k, \mu) = b\ b_{q} (1+\beta \mu^2) (1+\beta_{q} \mu^2) \ \tilde P(k),
\end{equation}
to obtain the parameters $b_{q}$ and $\beta_{q}$. 
%$\mu \equiv \hat{z}\cdot\hat{k}$:
For a given choice of $b$, $b_{q}$, $\beta$ and $\beta_{q}$, where $b_{q}$ and $\beta_{q}$ are related through $\beta_{q} = f/b_{q}$, we can calculate the theory-predicted multipoles of the auto-power spectrum, truncating at the hexadecapole, $\ell = 4$,
\begin{equation}
\tilde{P}_{\ell}(k) = b^2 C_{\ell}(\beta) \tilde{P}(k)
    \label{eq:p_ell}
\end{equation}
where
\iffalse
\begin{eqnarray}
C_{\ell}(\beta) &\equiv& \frac{2\ell+1}{2}\int_{-1}^{1}\left(1 + \beta\mu^2\right)^2 L_{\ell}d\mu \nonumber \\
&=&
\begin{cases}
    1 + \frac{2}{3}\beta + \frac{1}{5}\beta^2 & \ell = 0 \\
    \frac{4}{3}\beta + \frac{4}{7}\beta^2 & \ell = 2 \\
    \frac{8}{35}\beta^2 & \ell = 4
\end{cases} \; .
\label{eq:c_ell}
\end{eqnarray}
\fi
\begin{flalign}
C_{\ell}(\beta) &\equiv \frac{2\ell+1}{2}\int_{-1}^{1}\left(1 + \beta\mu^2\right)^2 L_{\ell}d\mu \nonumber \\
&=
\begin{cases}
    1 + \frac{2}{3}\beta + \frac{1}{5}\beta^2 & \ell = 0 \\
    \frac{4}{3}\beta + \frac{4}{7}\beta^2 & \ell = 2 \\
    \frac{8}{35}\beta^2 & \ell = 4
\end{cases} \; .
\label{eq:c_ell}
\end{flalign}
Similarly, we can express the cross-power spectrum with QSOs as:
\begin{equation}
    \tilde{P}_{q,\ell}(k) = b \ b_{q} C_{q,\ell}(\beta, \beta_{q}) \tilde{P}(k)
\end{equation}
with
\begin{eqnarray}
C_{q,\ell}(\beta) \equiv 
\begin{cases}
1 + \frac{1}{3}\beta + \frac{1}{3}\beta_{q} + \frac{1}{5}\beta \ \beta_{q} & \ell = 0 \\
\frac{2}{3}\beta + \frac{2}{3}\beta_{q} + \frac{4}{7}\beta \ \beta_{q} & \ell = 2 \\
\frac{8}{35}\beta \ \beta_{q} & \ell = 4
\end{cases} \; .
\label{eq:c_ellq}
\end{eqnarray}
%tracer bias and redshift-space distortion parameter  

Having obtained both the measured and the theoretical multipoles of the power spectrum, we can combine them into a single data vector, so as to supplement the poorly measured large scales ($k \lesssim 0.01 \hompc$) with the linear theory fit,
\begin{equation}
        P^{\rm comb}_\ell(k)=(1-w(k))P_\ell(k)+w(k)\,\tilde P_\ell(k),
\end{equation}
where $P_\ell$ and $\tilde P_\ell$ are the predictions from \textsc{AbacusSummit} and linear theory, respectively, and weighting function
\begin{equation}
    w(k)\equiv \frac{1}{2}\left[1-{\rm tanh}\left(\frac{k-k_{\rm pivot}}{\Delta k_w}\right)\right],
\end{equation}
which ensures smooth interpolation between the two limits. We use $\Delta k_w=0.01\hompc$, but manually fine-tune values of $k_{\rm pivot}$ for the different multipoles, based on their noisiness:
    \begin{align}
        &k_{{\rm pivot}, \ell=0}=0.03\hompc, \\
        &k_{{\rm pivot, \ell = 2}}=0.06\hompc,\hspace{12pt}
        k_{{\rm pivot, \ell = 4}}=0.09\hompc.
    \end{align}
{We have checked that the choice of a pivot scale for the monopole and quadrupole $\ell = 0, \ 2$ has negligible effect on the BAO feature. On the other hand, the hexadecapole, $\ell = 4$ is trickier to measure and hence a rather conservative scale cut is needed to ensure that the Hankel transform does not misbehave.}
      
Finally, we Hankel transform the power spectrum multipoles into correlation function multipoles, according to
\begin{equation}
    \xi_\ell(r) = \frac{i^{\ell}}{2\pi^2}\, \int_0^{\infty}\, k^2 j_{\ell}(k r)\,{P}^{\rm comb}_\ell(k)\,dk \; ,
\label{eq:xi_ell}
\end{equation}
where $j_{\ell}$ is the spherical Bessel function, and
\begin{equation}
    \xi_{q, \ell}(r) = \frac{i^{\ell}}{2\pi^2}\, \int_0^{\infty}\, k^2 j_{\ell}(k r)\,{P}^{\rm comb}_{q, \ell}(k)\,dk \; ,
\label{eq:xi_ellq}
\end{equation}
for the quasar cross-correlation. {The interpolation between theory and simulations ensures smooth integration and does not affect the clustering near the BAO scale and for smaller separations, $r$.}

\subsubsection{Comparing with linear and perturbation theory}
\label{sec:comp_lin}

One can model the effects of non-linear structure growth on the BAO feature via an anisotropic Gaussian smoothing of the linear power spectrum, effectively modifying Eq.~\ref{eq:lin} ~\citep{2007ApJ...664..660E}:
\begin{equation}
    \tilde{P}_{\rm nl}(k,\mu) = \exp[-k^2 \Sigma^2(\mu)/2]\cdot \tilde{P}(k)
    \label{eq:nl}
\end{equation}
where $\tilde{P}(k)$ is the linear power spectrum and
\begin{equation}
    \Sigma^2(\mu) = \mu^2 \Sigma_{\parallel}^2 + (1-\mu^2) \Sigma_{\perp}^2 \; ,
\end{equation}
where at redshift $z = 2.4$, we expect $\Sigma_{\parallel} \simeq 6.41 \mpcoh$ and $\Sigma_{\perp} \simeq 3.26 \mpcoh$. In principle, this equation implies that $\tilde{P}_{\ell}(k)$, Eq.~\ref{eq:p_ell}, cannot be decomposed into a $\beta$- and a $k$-dependent factor, but instead that the integrals should be re-evaluated for each value of $\beta$. However, since $\Sigma \simeq 5\mpcoh$ is expected to be smaller than the peak full-width half-maximum, following \citet{2013JCAP...03..024K}, we adopt the approximation:
\begin{equation}
\label{eq:bao}
    P_{{\rm nl}, \ell}(k) \simeq \exp(-k^2\Sigma_{\ell}^2(\beta)/2) \cdot \tilde{P}_{\ell}(k),
\end{equation}
so that each multipole undergoes a different amount of isotropic broadening according to:
\begin{equation}
    \Sigma^2_{\ell}(\beta) \equiv f_{\ell}(\beta)\cdot\Sigma_{\parallel}^2 + (1 - f_{\ell}(\beta))\cdot\Sigma_{\perp}^2
\end{equation}
where
\begin{equation}
    f_{\ell}(\beta) \equiv \frac{\int_{-1}^{+1}\,\mu^2 \left(1 + \beta\mu^2\right)^2 L_{\ell}(\mu)\,d\mu}
    {\int_{-1}^{+1}\,\left(1 + \beta\mu^2\right)^2 L_{\ell}(\mu)\,d\mu} =
\begin{cases}
\frac{35+42\beta+15\beta^2}{105+70\beta+21\beta^2} & \ell = 0 \\
\frac{7+12\beta+5\beta^2}{14\beta+6\beta^2} & \ell = 2 \\
\frac{15}{11}+\frac{2}{\beta} & \ell = 4
\end{cases} \; .
\end{equation}
Similarly, we can perform the analogous integral to arrive at the equations predicting the non-linear broadening in the Ly$\alpha$-QSO correlation function:
\begin{equation}
f_{q, \ell}(\beta) \equiv 
\begin{cases}
\frac{35+21\beta+21\beta_{q}+15\beta \beta_{q}}{105+35\beta+35\beta_{q}+21\beta \beta_{q}} & \ell = 0 \\
\frac{7+6\beta+6\beta_{q}+5\beta \beta_{q}}{7\beta+7\beta_{q}+6\beta \beta_{q}} & \ell = 2 \\
\frac{15}{11}+\frac{1}{\beta}+\frac{1}{\beta_{q}} & \ell = 4
\end{cases} \; .
\end{equation}
Next, we study the correlation functions computed from our $N$-body mocks and compare them with linear and perturbation theory, with the latter following the derived form above.

\begin{figure*}
    \centering
    \includegraphics[width=0.98\textwidth]{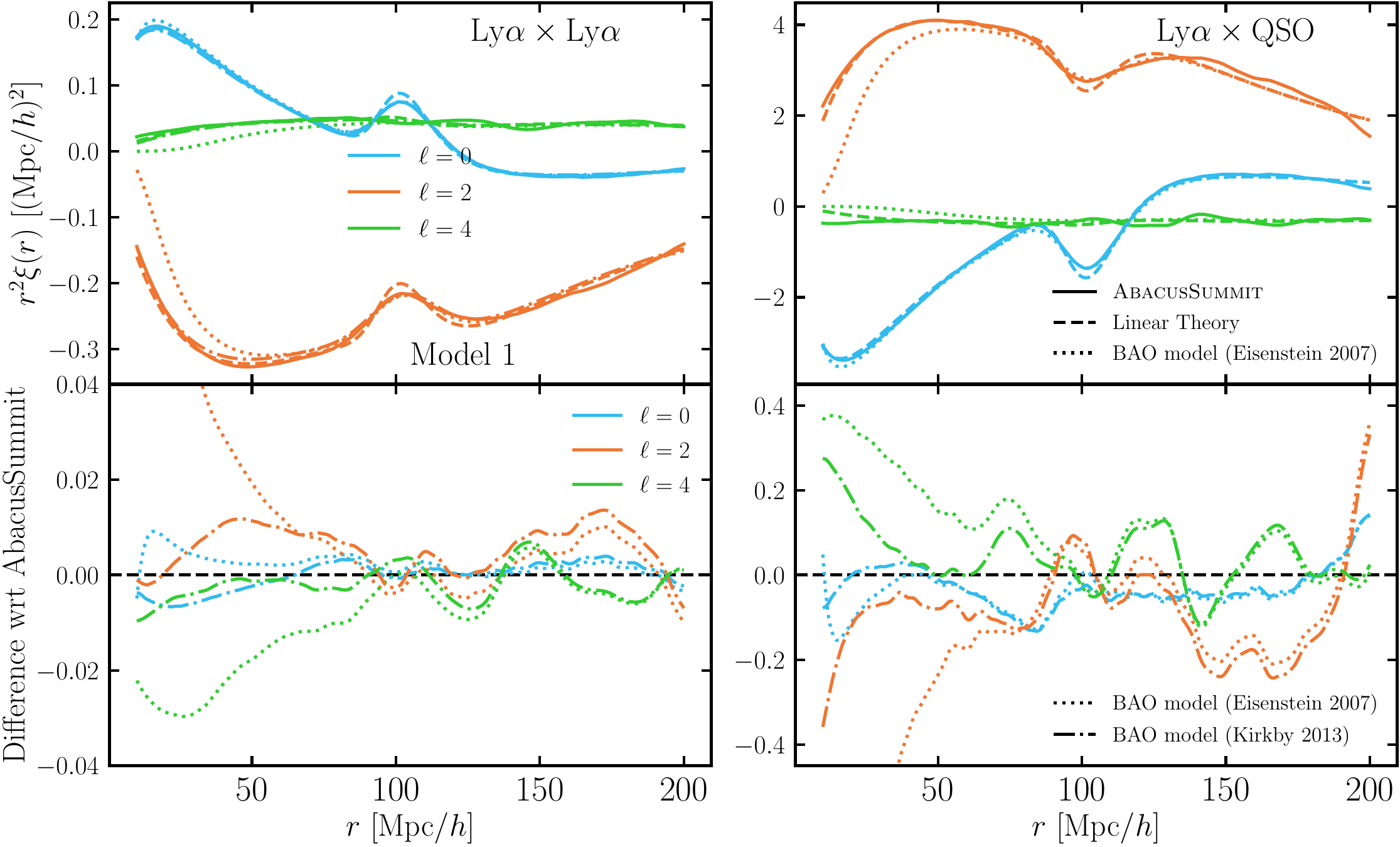}
    \caption{{Multipoles of the Ly$\alpha$-Ly$\alpha$ and Ly$\alpha$-QSO correlation function, $\xi_\ell$, comparing measurements from our \lyaf\ mocks on \textsc{AbacusSummit} for Model 1 (see Table~\ref{tab:models}) with the theoretical prediction from linear theory and the BAO broadening (LPT-based) models of \citet{2007ApJ...664..660E} and \citet{2013JCAP...03..024K}. Difference with respect to the simulations is shown in the bottom. We find clear evidence for the broadening of the BAO peak in our simulations and excellent agreement with the LPT-based models. Note that the \citet{2007ApJ...664..660E} is not suitable on small scales, i.e., below $r \lesssim 80\mpcoh$, as it overly suppresses the power. On small scales below $r \lesssim 30\mpcoh$, the \citet{2013JCAP...03..024K} BAO model prediction of the cross-correlation with quasars (right panel) deviates from the simulation, which we can attribute to the breakdown of the Kaiser approximation and small-scale physics effects (such as non-linear and Fingers-of-God effects).}} 
    \label{fig:multipole_theory}
\end{figure*}
%{2007ApJ...664..660E} {2013JCAP...03..024K} 2014MNRAS.439...83A,

In Fig.~\ref{fig:multipole_theory}, we show the correlation function of the Ly$\alpha$-Ly$\alpha$ and Ly$\alpha$-QSO tracers for \textsc{AbacusSummit}, linear theory (Eq.~\ref{eq:p_ell}) and {two BAO models based on Lagrangian Perturbation Theory (LPT): 
{1) \citet{2007ApJ...664..660E}, which applies the Gaussian smoothing to the entire linear power (Eq.~\ref{eq:nl}); 
2) \citet{2013JCAP...03..024K}, which decomposes the power spectrum into a `wiggle' and `no-wiggle' component, and only applies the Gaussian smoothing to the peak component.}
%1) \citet{2007ApJ...664..660E}, which uses Eq.~\ref{eq:bao} and transforms it to $\xi_\ell$ via Eq.~\ref{eq:xi_ell}; 2) \citet{2013JCAP...03..024K}, which splits the power spectrum into a `wiggle' and `no-wiggle' component. 
We refer to these two models as LPT-based, but stress that they do not adopt the full LPT toolkit to model small scales, but instead offer BAO scale correction to recover the non-linear broadening of the BAO peak.} The \textsc{AbacusSummit} measurements are obtained by combining all six boxes for one of the four models (in particular, Model 1; see Table~\ref{tab:models}), performing a Hankel transform and averaging over them, to get smoother behavior. The BAO feature is visible in all curves except for $\ell = 4$, which is both noisier and has a weaker BAO signal. We note that the perturbation theory curve for $\ell = 4$ is also lacking a visible peak. It is clear that the sharpness of the linear theory prediction is substantially suppressed in the simulation, providing strong evidence of non-linear broadening. It is further reassuring that the perturbation theory predictions are in good agreement with the simulation at the BAO scale, especially for $\ell = 0$ and $\ell = 2$. 

{On scales smaller than $r \lesssim 80\mpcoh$, the simplified perturbation theory model of \citet{2007ApJ...664..660E} is inaccurate, as it over-smooths the power spectrum on small scales. Models that only smooth the peak component \citep{2013JCAP...03..024K} can be trusted down to smaller scales. Remaining differences between the simulations and the \citet{2013JCAP...03..024K} model on small scales, i.e. $r \lesssim 30\mpcoh$, can be attributed to the various simplifications of the Kaiser model, which neglects non-linear effects. Nonetheless, these appear to be quite small for the \lya\ auto-correlation function, indicating that the Kaiser approximation works surprisingly well in that regime. They are, however, more pronounced when studying cross-correlations with the QSOs, suggesting that the QSO field is more affected by non-linear effects, as one would expect. One of the main uses of our mocks will be to test the scales at which the Kaiser approximation breaks down, as this is a central question for the analysis pipelines being developed.}

\begin{figure*}
    \centering
    \includegraphics[width=0.98\textwidth]{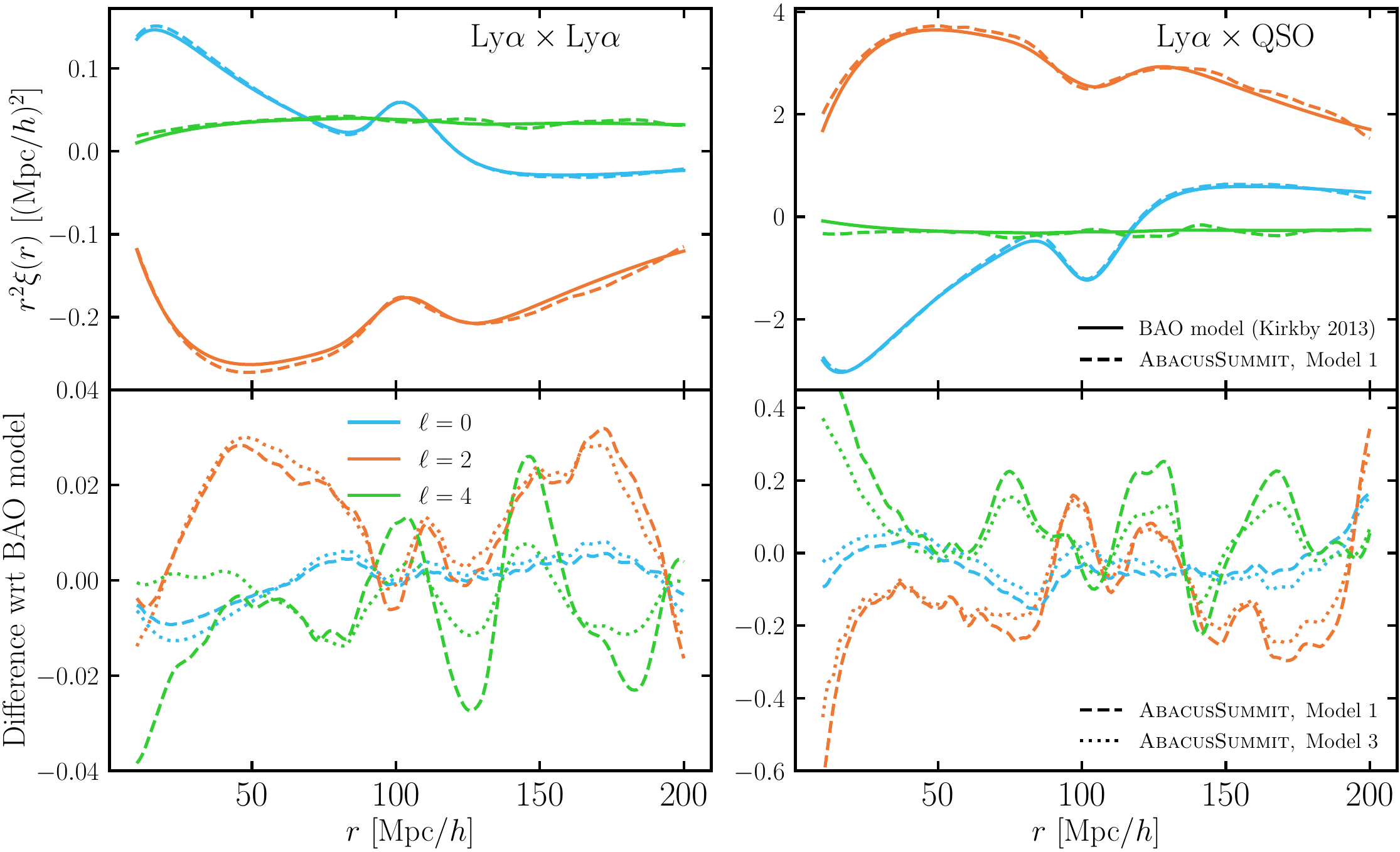}
    \caption{{Multipoles of the Ly$\alpha$-Ly$\alpha$ and Ly$\alpha$-QSO correlation function, $\xi_\ell$, comparing measurements from our \lyaf\ mocks on \textsc{AbacusSummit} for Model 1 and Model 3 (see Table~\ref{tab:models}) with the theoretical prediction from the BAO broadening (LPT-based) model of \citet{2013JCAP...03..024K}. The bottom panels show the difference between the two mock measurements and the \citet{2013JCAP...03..024K} BAO model, while the top shows Model 1 and the \citet{2013JCAP...03..024K} prediction matching the bias and $\beta$ values. Reassuringly, the $\ell = 0$ and $\ell = 2$ multipoles of Models 1 and 3 are very consistent with each other, suggesting that the \lya\ painting technique has little effect on intermediate and large scales. Larger differences between the two are seen for the $\ell = 4$ case, which needs to be studied in more details. As in Fig.~\ref{fig:multipole_theory}, the Kaiser approximation provides a poor match below $r \lesssim 30\mpcoh$ for the cross-correlation, indicating that more careful modeling needs to be done to achieve sufficient precision. Note that we have rescaled the difference curves (bottom panels) for Model 1 by the pre-factors $C_{\ell}(\beta)$ and $C_{q,\ell}(\beta)$ (see Eq.~\ref{eq:c_ell} and Eq.~\ref{eq:c_ellq}) to account to linear order for the different values of $\beta$.}} 
    \label{fig:multipole_eboss}
\end{figure*}

{In Fig.~\ref{fig:multipole_eboss}, we explore how the broadening changes for two of the four different models we have adopted in generating the \lyaf\ mocks, namely, Model 1 and Model 3. We find that this choice has little to no effect on the broadened BAO feature, and thus the comparison with perturbation theory remains qualitatively unchanged. Furthermore, the $\ell = 0$ and $\ell =2 $ multipoles of Models 1 and 3 are very consistent with each other across a wide range of scales, suggesting that the \lya\ painting technique hardly affects these multipoles. Larger differences are seen for the $\ell = 4$ case, which is noisier and hence more difficult to measure, so we leave a more detailed study for the future. As in Fig.~\ref{fig:multipole_theory}, the Kaiser approximation of the BAO model provides a poor match below $r \lesssim 30\mpcoh$, indicating that the cross-correlations may need to be modeled beyond the Kaiser approximation with non-linear effects properly accounted for. Note that when showing the difference curves, we have rescaled the Model 1 multipoles by the pre-factors $C_{\ell}(\beta)$ and $C_{q,\ell}(\beta)$ (see Eq.~\ref{eq:c_ell} and Eq.~\ref{eq:c_ellq}) to account to linear order for the different values of $\beta$ (see Table~\ref{tab:models}) and make the comparison with Model 3 more straightforward to see.}

\section{Summary}
\label{sec:conc}
The absorption of \Lya\   photons by hydrogen clouds imparts a characteristic signature on the spectra of high-redshift sources, known as the \lyaf. These features, revealing the cosmic web of filamentary structures, have become a powerful tool for the study of large-scale structure in observational cosmology through measurements of their power spectrum and clustering. Accurately measuring these requires careful accounting of the systematic errors and is essential in order to extract cosmological constraints. The only reliable way of doing this is to generate random realizations of multiple \Lya\  absorption spectra in a survey and decorate them with various systematic effects so as to obtain a maximum realism data set. Examples of such systematics include a thorough modeling of the quasar continuum, which is used to infer the transmitted flux fraction, a modeling of the variable spectral resolution and noise, a calibration of the flux errors, an evaluation of the impact of redshift evolution, Damped Lyman alpha systems (DLAs), Lyman limit systems (LLS), metal absorption lines, and the cosmic ionizing background. The mock surveys needed to investigate these questions must include a large number of lines-of-sight over a large volume so as to satisfy the ambitious requirements set by surveys such as DESI, while also including small-scale fluctuations which contain a lot of valuable information through redshift-space distortions and the suppression of the power spectrum. Needless to say, this is extremely computationally challenging, and cosmologists typically resolve to having two sets of simulations for modeling the large- and small-scale observables. While our mocks do not provide accuracy down to the smallest scales needed to constrain neutrino and dark matter models, they present a first step to reconciling the large scales necessary for a BAO peak study and the small scales used to extract structure growth rate information in a single suite of mock catalogs.

Our \lyaf\ synthetic catalogs are generated on the largest $N$-body simulation suite \textsc{AbacusSummit} and are publicly available for six of the \texttt{base} boxes, $L_{\rm box} = 2\gpcoh$, at the fiducial \textit{Planck} 2018 cosmology. Mock skewers are available on a regular grid with $6912^3$ cells, and we output four different versions of our recipe per each observer location (at infinity along the $z$ axis and along the $y$ axis). In particular, we utilize the Fluctuating Gunn-Peterson Approximation (FGPA) and a modification thereof to transform the dark matter density field into a \lyaf\ catalog (see Section~\ref{sec:mocks} for details on our methods) and aim to match various \Lya\   observables extracted from a hydrodynamical simulation. Namely, we employ the high-realism \lyaf\ skewers produced by \citet{2022ApJ...930..109Q} for the hydro run TNG300-1 and calibrate our FGPA-based mocks against the mean and standard deviation of the transmission flux as well as the 1D and 3D \Lya\   power spectrum. We make the prime choices for our model parameters through the comparison between TNG300-1 and its low-resolution dark-matter-only counterpart TNG300-3-DM and find that our simplistic recipe yields a satisfactory agreement between the power spectra, as presented in Fig.~\ref{fig:power_tng}. We then go on to apply this prescription to the \textsc{AbacusSummit} boxes, which have similar resolution to TNG300-3-DM, finding that the level of agreement is retained and the largest scales reachable largely extended by an order of magnitude (see Fig.~\ref{fig:power_abacus}). Next, we study the correlation function multipoles of \Lya-\Lya\  and \Lya-QSO in Fig.~\ref{fig:multipole_theory}, which demonstrates for the first time in \Lya\   simulations the effect of non-linear clustering on the BAO peak. 
%We also compare these multipoles with the eBOSS DR16 predictions in Fig.~\ref{fig:multipole_eboss}, finding that the correlation function shapes from our mocks match reasonably well observations. 
We find differences on small scales between the linear model (i.e., Kaiser approximation) and our mocks, especially in cross-correlations with the QSO population, which would be important to account for in the analysis of \lyaf\ data.

Apart from being useful for testing systematics and the analysis pipeline, our mocks also open the doors for modeling novel statistics and joint probes with other tracers. As an example, developing and testing summary statistics that maximally use the 3D information in the \lyaf\ such as the $P_\times$ estimator of \citet{2018JCAP...01..003F} would be crucial to fully realizing the potential of the \Lya\   probe. The question of whether the \lyaf\ measurements from current surveys such as DESI can be utilized to constrain the growth rate $f$ (e.g., as done in 3$\times$2-pt \Lya-QSO analysis), is also not yet fully resolved. By grafting the survey properties onto our mocks and performing the analysis on them as if on real data, we can tackle this problem and quote forecasts for the expected constraining power. It is also essential that we understand the scales at which linear theory breaks from our mocks and develop theoretical models that can recover the small-scale clustering correctly. Finally, a particularly exciting venue to explore is the development of combined analysis tools for \lyaf\ and CMB lensing, which promises to break important degeneracies in our models (such as the two bias parameters characteristic of \Lya\  observables). As an initial step in near-term work, we plan to develop a model able to reproduce the joint data vector, using already available CMB and light cone products \citep{2022MNRAS.509.2194H,Hadzhiyska+2023}, and successfully glean cosmological information from it.

{Another important direction in which we could further develop our mocks is by adding realistic observational effects that are otherwise difficult to study analytically. These include (but are not limited to) the effects of Damped Lyman alpha systems (DLAs), Lyman limit systems (LLS), metal absorption lines, and the cosmic ionizing (UVB) background, which could be added to our mocks following prescriptions similar to the FGPA method employed in this work.} In later versions of our mocks, we also plan to adopt density estimation techniques better suited for low-density regions such as phase-space tessellation, improved velocity field estimation schemes, and machine-learning methods for painting hydro simulation results on $N$-body simulations, which would improve the small-scale synthetic absorption signal. {That way, we also hope to make our model more flexible to matching the observed power spectrum and correlation function across a wider range of scales and with a higher level of accuracy.} In addition, we have planned to run our model on the light cone, so as to enable maximum realism including redshift evolution and curved sky effects, as well as facilitate joint studies with other tracers such as the CMB and weak lensing. The \lyaf\ is still a largely unexplored resource that is brimming with astrophysical and cosmological information, waiting to be relinquished and utilized to uncover fundamental truths about our Universe.

\section*{Acknowledgements}

We would like to thank Mahdi Qezlou, Vid Ir\v{s}i\v{c}, Patrick McDonald, John Peacock, An\v{z}e Slosar, Julien Guy, Zarija Luki\'c and Martin White for illuminating discussions over the course of this project. %In addition, we are grateful to  for helpful comments and suggestions.
AFR acknowledges support from the Spanish Ministry of Science and Innovation through the program Ramon y Cajal (RYC-2018-025210) and from the European Union’s Horizon Europe research and innovation programme (COSMO-LYA, grant agreement 101044612). IFAE is partially funded by the CERCA program of the Generalitat de Catalunya. AC acknowledges support from the United States Department of Energy, Office of High Energy Physics under Award Number DE-SC-0011726.

This research is supported by the Director, Office of Science, Office of High Energy Physics of the U.S. Department of Energy under Contract No. DE–AC02–05CH11231, and by the National Energy Research Scientific Computing Center, a DOE Office of Science User Facility under the same contract; additional support for DESI is provided by the U.S. National Science Foundation, Division of Astronomical Sciences under Contract No. AST-0950945 to the NSF’s National Optical-Infrared Astronomy Research Laboratory; the Science and Technologies Facilities Council of the United Kingdom; the Gordon and Betty Moore Foundation; the Heising-Simons Foundation; the French Alternative Energies and Atomic Energy Commission (CEA); the National Council of Science and Technology of Mexico (CONACYT); the Ministry of Science and Innovation of Spain (MICINN), and by the DESI Member Institutions: \url{https://www.desi.lbl.gov/collaborating-institutions}.

The authors are honored to be permitted to conduct scientific research on Iolkam Du’ag (Kitt Peak), a mountain with particular significance to the Tohono O’odham Nation.

%%%%%%%%%%%%%%%%%%%%%%%%%%%%%%%%%%%%%%%%%%%%%%%%%%
\section*{Data Availability}

We make all our synthetic maps and catalogues publicly available on Globus through NERSC SHARE at this link: \url{https://app.globus.org/file-manager?origin_id=9ce29982-eed1-11ed-9bb4-c9bb788c490e&path=\%2F} under the name ``AbacusSummit Lyman Alpha Forest''. Data points for the figures are available at \url{https://doi.org/10.5281/zenodo.7926520}.



\bibliographystyle{mnras}
\bibliography{example} % if your bibtex file is called example.bib

\begin{thebibliography}{}
\makeatletter
\relax
\def\mn@urlcharsother{\let\do\@makeother \do\$\do\&\do\#\do\^\do\_\do\%\do\~}
\def\mn@doi{\begingroup\mn@urlcharsother \@ifnextchar [ {\mn@doi@}
  {\mn@doi@[]}}
\def\mn@doi@[#1]#2{\def\@tempa{#1}\ifx\@tempa\@empty \href
  {http://dx.doi.org/#2} {doi:#2}\else \href {http://dx.doi.org/#2} {#1}\fi
  \endgroup}
\def\mn@eprint#1#2{\mn@eprint@#1:#2::\@nil}
\def\mn@eprint@arXiv#1{\href {http://arxiv.org/abs/#1} {{\tt arXiv:#1}}}
\def\mn@eprint@dblp#1{\href {http://dblp.uni-trier.de/rec/bibtex/#1.xml}
  {dblp:#1}}
\def\mn@eprint@#1:#2:#3:#4\@nil{\def\@tempa {#1}\def\@tempb {#2}\def\@tempc
  {#3}\ifx \@tempc \@empty \let \@tempc \@tempb \let \@tempb \@tempa \fi \ifx
  \@tempb \@empty \def\@tempb {arXiv}\fi \@ifundefined
  {mn@eprint@\@tempb}{\@tempb:\@tempc}{\expandafter \expandafter \csname
  mn@eprint@\@tempb\endcsname \expandafter{\@tempc}}}

\bibitem[\protect\citeauthoryear{{Abbott} et~al.,}{{Abbott}
  et~al.}{2018}]{2018PhRvD..98d3526A}
{Abbott} T.~M.~C.,  et~al., 2018, \mn@doi [\prd] {10.1103/PhysRevD.98.043526},
  \href {https://ui.adsabs.harvard.edu/abs/2018PhRvD..98d3526A} {98, 043526}

\bibitem[\protect\citeauthoryear{{Abel}, {Hahn}  \& {Kaehler}}{{Abel}
  et~al.}{2012}]{2012MNRAS.427...61A}
{Abel} T.,  {Hahn} O.,   {Kaehler} R.,  2012, \mn@doi [\mnras]
  {10.1111/j.1365-2966.2012.21754.x}, \href
  {https://ui.adsabs.harvard.edu/abs/2012MNRAS.427...61A} {427, 61}

\bibitem[\protect\citeauthoryear{{Arinyo-i-Prats}, {Miralda-Escud{\'e}}, {Viel}
   \& {Cen}}{{Arinyo-i-Prats} et~al.}{2015}]{2015JCAP...12..017A}
{Arinyo-i-Prats} A.,  {Miralda-Escud{\'e}} J.,  {Viel} M.,   {Cen} R.,  2015,
  \mn@doi [\jcap] {10.1088/1475-7516/2015/12/017}, \href
  {https://ui.adsabs.harvard.edu/abs/2015JCAP...12..017A} {2015, 017}

\bibitem[\protect\citeauthoryear{{Ata} et~al.,}{{Ata}
  et~al.}{2018}]{Ata:2018MNRAS.473.4773A}
{Ata} M.,  et~al., 2018, \mn@doi [\mnras] {10.1093/mnras/stx2630}, \href
  {https://ui.adsabs.harvard.edu/abs/2018MNRAS.473.4773A} {473, 4773}

\bibitem[\protect\citeauthoryear{{Baur}, {Palanque-Delabrouille}, {Y{\`e}che},
  {Boyarsky}, {Ruchayskiy}, {Armengaud}  \& {Lesgourgues}}{{Baur}
  et~al.}{2017}]{2017JCAP...12..013B}
{Baur} J.,  {Palanque-Delabrouille} N.,  {Y{\`e}che} C.,  {Boyarsky} A.,
  {Ruchayskiy} O.,  {Armengaud} {\'E}.,   {Lesgourgues} J.,  2017, \mn@doi
  [\jcap] {10.1088/1475-7516/2017/12/013}, \href
  {https://ui.adsabs.harvard.edu/abs/2017JCAP...12..013B} {2017, 013}

\bibitem[\protect\citeauthoryear{{Bautista} et~al.,}{{Bautista}
  et~al.}{2015}]{Bautista:2015JCAP...05..060B}
{Bautista} J.~E.,  et~al., 2015, \mn@doi [\jcap]
  {10.1088/1475-7516/2015/05/060}, \href
  {https://ui.adsabs.harvard.edu/abs/2015JCAP...05..060B} {2015, 060}

\bibitem[\protect\citeauthoryear{{Bernardeau}, {Colombi}, {Gazta{\~n}aga}  \&
  {Scoccimarro}}{{Bernardeau} et~al.}{2002}]{Bernardeau:2002PhR...367....1B}
{Bernardeau} F.,  {Colombi} S.,  {Gazta{\~n}aga} E.,   {Scoccimarro} R.,  2002,
  \mn@doi [\physrep] {10.1016/S0370-1573(02)00135-7}, \href
  {https://ui.adsabs.harvard.edu/abs/2002PhR...367....1B} {367, 1}

\bibitem[\protect\citeauthoryear{{Bi} \& {Davidsen}}{{Bi} \&
  {Davidsen}}{1997}]{Bi:1997ApJ...479..523B}
{Bi} H.,  {Davidsen} A.~F.,  1997, \mn@doi [\apj] {10.1086/303908}, \href
  {https://ui.adsabs.harvard.edu/abs/1997ApJ...479..523B} {479, 523}

\bibitem[\protect\citeauthoryear{{Bi}, {Boerner}  \& {Chu}}{{Bi}
  et~al.}{1992}]{Bi:1992A&A...266....1B}
{Bi} H.~G.,  {Boerner} G.,   {Chu} Y.,  1992, \aap, \href
  {https://ui.adsabs.harvard.edu/abs/1992A&A...266....1B} {266, 1}

\bibitem[\protect\citeauthoryear{{Bird}}{{Bird}}{2017}]{2017ascl.soft10012B}
{Bird} S.,  2017, {FSFE: Fake Spectra Flux Extractor}, Astrophysics Source Code
  Library, record ascl:1710.012 (\mn@eprint {ascl} {1710.012})

\bibitem[\protect\citeauthoryear{{Bird}, {Peiris}, {Viel}  \& {Verde}}{{Bird}
  et~al.}{2011}]{Bird2011}
{Bird} S.,  {Peiris} H.~V.,  {Viel} M.,   {Verde} L.,  2011, \mn@doi [\mnras]
  {10.1111/j.1365-2966.2011.18245.x}, \href
  {https://ui.adsabs.harvard.edu/abs/2011MNRAS.413.1717B} {413, 1717}

\bibitem[\protect\citeauthoryear{{Bird}, {Haehnelt}, {Neeleman}, {Genel},
  {Vogelsberger}  \& {Hernquist}}{{Bird} et~al.}{2015}]{2015MNRAS.447.1834B}
{Bird} S.,  {Haehnelt} M.,  {Neeleman} M.,  {Genel} S.,  {Vogelsberger} M.,
  {Hernquist} L.,  2015, \mn@doi [\mnras] {10.1093/mnras/stu2542}, \href
  {https://ui.adsabs.harvard.edu/abs/2015MNRAS.447.1834B} {447, 1834}

\bibitem[\protect\citeauthoryear{{Chabanier} et~al.,}{{Chabanier}
  et~al.}{2019}]{Chabanier2019}
{Chabanier} S.,  et~al., 2019, \mn@doi [\jcap] {10.1088/1475-7516/2019/07/017},
  \href {https://ui.adsabs.harvard.edu/abs/2019JCAP...07..017C} {2019, 017}

\bibitem[\protect\citeauthoryear{{Chabanier} et~al.,}{{Chabanier}
  et~al.}{2023}]{2023MNRAS.518.3754C}
{Chabanier} S.,  et~al., 2023, \mn@doi [\mnras] {10.1093/mnras/stac3294}, \href
  {https://ui.adsabs.harvard.edu/abs/2023MNRAS.518.3754C} {518, 3754}

\bibitem[\protect\citeauthoryear{{Chaussidon} et~al.,}{{Chaussidon}
  et~al.}{2023}]{2023ApJ...944..107C}
{Chaussidon} E.,  et~al., 2023, \mn@doi [\apj] {10.3847/1538-4357/acb3c2},
  \href {https://ui.adsabs.harvard.edu/abs/2023ApJ...944..107C} {944, 107}

\bibitem[\protect\citeauthoryear{{Cole} et~al.,}{{Cole}
  et~al.}{2005}]{Cole:2005MNRAS.362..505C}
{Cole} S.,  et~al., 2005, \mn@doi [\mnras] {10.1111/j.1365-2966.2005.09318.x},
  \href {https://ui.adsabs.harvard.edu/abs/2005MNRAS.362..505C} {362, 505}

\bibitem[\protect\citeauthoryear{{Coles} \& {Jones}}{{Coles} \&
  {Jones}}{1991}]{Coles:1991MNRAS.248....1C}
{Coles} P.,  {Jones} B.,  1991, \mn@doi [\mnras] {10.1093/mnras/248.1.1}, \href
  {https://ui.adsabs.harvard.edu/abs/1991MNRAS.248....1C} {248, 1}

\bibitem[\protect\citeauthoryear{{Croft}, {Weinberg}, {Katz}  \&
  {Hernquist}}{{Croft} et~al.}{1998a}]{Croft1998}
{Croft} R. A.~C.,  {Weinberg} D.~H.,  {Katz} N.,   {Hernquist} L.,  1998a,
  \mn@doi [\apj] {10.1086/305289}, \href
  {https://ui.adsabs.harvard.edu/abs/1998ApJ...495...44C} {495, 44}

\bibitem[\protect\citeauthoryear{{Croft}, {Weinberg}, {Katz}  \&
  {Hernquist}}{{Croft} et~al.}{1998b}]{Croft:1998ApJ...495...44C}
{Croft} R. A.~C.,  {Weinberg} D.~H.,  {Katz} N.,   {Hernquist} L.,  1998b,
  \mn@doi [\apj] {10.1086/305289}, \href
  {https://ui.adsabs.harvard.edu/abs/1998ApJ...495...44C} {495, 44}

\bibitem[\protect\citeauthoryear{{Croft}, {Weinberg}, {Pettini}, {Hernquist}
  \& {Katz}}{{Croft} et~al.}{1999}]{Croft1999}
{Croft} R. A.~C.,  {Weinberg} D.~H.,  {Pettini} M.,  {Hernquist} L.,   {Katz}
  N.,  1999, \mn@doi [\apj] {10.1086/307438}, \href
  {https://ui.adsabs.harvard.edu/abs/1999ApJ...520....1C} {520, 1}

\bibitem[\protect\citeauthoryear{{Croft}, {Weinberg}, {Bolte}, {Burles},
  {Hernquist}, {Katz}, {Kirkman}  \& {Tytler}}{{Croft}
  et~al.}{2002}]{Croft2002}
{Croft} R. A.~C.,  {Weinberg} D.~H.,  {Bolte} M.,  {Burles} S.,  {Hernquist}
  L.,  {Katz} N.,  {Kirkman} D.,   {Tytler} D.,  2002, \mn@doi [\apj]
  {10.1086/344099}, \href
  {https://ui.adsabs.harvard.edu/abs/2002ApJ...581...20C} {581, 20}

\bibitem[\protect\citeauthoryear{{Cuceu}, {Font-Ribera}, {Joachimi}  \&
  {Nadathur}}{{Cuceu} et~al.}{2021}]{2021MNRAS.506.5439C}
{Cuceu} A.,  {Font-Ribera} A.,  {Joachimi} B.,   {Nadathur} S.,  2021, \mn@doi
  [\mnras] {10.1093/mnras/stab1999}, \href
  {https://ui.adsabs.harvard.edu/abs/2021MNRAS.506.5439C} {506, 5439}

\bibitem[\protect\citeauthoryear{{Cuceu} et~al.,}{{Cuceu}
  et~al.}{2022a}]{2022arXiv220912931C}
{Cuceu} A.,  et~al., 2022a, arXiv e-prints, \href
  {https://ui.adsabs.harvard.edu/abs/2022arXiv220912931C} {p. arXiv:2209.12931}

\bibitem[\protect\citeauthoryear{{Cuceu}, {Font-Ribera}, {Nadathur}, {Joachimi}
   \& {Martini}}{{Cuceu} et~al.}{2022b}]{2022arXiv220913942C}
{Cuceu} A.,  {Font-Ribera} A.,  {Nadathur} S.,  {Joachimi} B.,   {Martini} P.,
  2022b, arXiv e-prints, \href
  {https://ui.adsabs.harvard.edu/abs/2022arXiv220913942C} {p. arXiv:2209.13942}

\bibitem[\protect\citeauthoryear{{DESI Collaboration} et~al.,}{{DESI
  Collaboration} et~al.}{2016a}]{2016arXiv161100036D}
{DESI Collaboration} et~al., 2016a, arXiv e-prints, \href
  {https://ui.adsabs.harvard.edu/abs/2016arXiv161100036D} {p. arXiv:1611.00036}

\bibitem[\protect\citeauthoryear{{DESI Collaboration} et~al.,}{{DESI
  Collaboration} et~al.}{2016b}]{2016arXiv161100037D}
{DESI Collaboration} et~al., 2016b, arXiv e-prints, \href
  {https://ui.adsabs.harvard.edu/abs/2016arXiv161100037D} {p. arXiv:1611.00037}

\bibitem[\protect\citeauthoryear{{DESI Collaboration} et~al.,}{{DESI
  Collaboration} et~al.}{2022}]{2022arXiv220510939A}
{DESI Collaboration} et~al., 2022, arXiv e-prints, \href
  {https://ui.adsabs.harvard.edu/abs/2022arXiv220510939A} {p. arXiv:2205.10939}

\bibitem[\protect\citeauthoryear{{Dark Energy Survey Collaboration}
  et~al.,}{{Dark Energy Survey Collaboration}
  et~al.}{2016}]{2016MNRAS.460.1270D}
{Dark Energy Survey Collaboration} et~al., 2016, \mn@doi [\mnras]
  {10.1093/mnras/stw641}, \href
  {https://ui.adsabs.harvard.edu/abs/2016MNRAS.460.1270D} {460, 1270}

\bibitem[\protect\citeauthoryear{{Eisenstein} et~al.,}{{Eisenstein}
  et~al.}{2005}]{Eisenstein:2005ApJ...633..560E}
{Eisenstein} D.~J.,  et~al., 2005, \mn@doi [\apj] {10.1086/466512}, \href
  {https://ui.adsabs.harvard.edu/abs/2005ApJ...633..560E} {633, 560}

\bibitem[\protect\citeauthoryear{{Eisenstein}, {Seo}  \& {White}}{{Eisenstein}
  et~al.}{2007}]{2007ApJ...664..660E}
{Eisenstein} D.~J.,  {Seo} H.-J.,   {White} M.,  2007, \mn@doi [\apj]
  {10.1086/518755}, \href
  {https://ui.adsabs.harvard.edu/abs/2007ApJ...664..660E} {664, 660}

\bibitem[\protect\citeauthoryear{{Farr} et~al.,}{{Farr}
  et~al.}{2020}]{2020JCAP...03..068F}
{Farr} J.,  et~al., 2020, \mn@doi [\jcap] {10.1088/1475-7516/2020/03/068},
  \href {https://ui.adsabs.harvard.edu/abs/2020JCAP...03..068F} {2020, 068}

\bibitem[\protect\citeauthoryear{{Faucher-Gigu{\`e}re}, {Lidz}, {Hernquist}  \&
  {Zaldarriaga}}{{Faucher-Gigu{\`e}re} et~al.}{2008}]{2008ApJ...688...85F}
{Faucher-Gigu{\`e}re} C.-A.,  {Lidz} A.,  {Hernquist} L.,   {Zaldarriaga} M.,
  2008, \mn@doi [\apj] {10.1086/592289}, \href
  {https://ui.adsabs.harvard.edu/abs/2008ApJ...688...85F} {688, 85}

\bibitem[\protect\citeauthoryear{{Flaugher} et~al.,}{{Flaugher}
  et~al.}{2015}]{2015AJ....150..150F}
{Flaugher} B.,  et~al., 2015, \mn@doi [\aj] {10.1088/0004-6256/150/5/150},
  \href {https://ui.adsabs.harvard.edu/abs/2015AJ....150..150F} {150, 150}

\bibitem[\protect\citeauthoryear{{Font-Ribera}, {McDonald}  \&
  {Miralda-Escud{\'e}}}{{Font-Ribera}
  et~al.}{2012}]{Font-Ribera:2012JCAP...01..001F}
{Font-Ribera} A.,  {McDonald} P.,   {Miralda-Escud{\'e}} J.,  2012, \mn@doi
  [\jcap] {10.1088/1475-7516/2012/01/001}, \href
  {https://ui.adsabs.harvard.edu/abs/2012JCAP...01..001F} {2012, 001}

\bibitem[\protect\citeauthoryear{{Font-Ribera}, {McDonald}  \&
  {Slosar}}{{Font-Ribera} et~al.}{2018}]{2018JCAP...01..003F}
{Font-Ribera} A.,  {McDonald} P.,   {Slosar} A.,  2018, \mn@doi [\jcap]
  {10.1088/1475-7516/2018/01/003}, \href
  {https://ui.adsabs.harvard.edu/abs/2018JCAP...01..003F} {2018, 003}

\bibitem[\protect\citeauthoryear{{Garrison}, {Eisenstein}  \&
  {Pinto}}{{Garrison} et~al.}{2019}]{2019MNRAS.485.3370G}
{Garrison} L.~H.,  {Eisenstein} D.~J.,   {Pinto} P.~A.,  2019, \mn@doi [\mnras]
  {10.1093/mnras/stz634}, \href
  {https://ui.adsabs.harvard.edu/abs/2019MNRAS.485.3370G} {485, 3370}

\bibitem[\protect\citeauthoryear{{Garrison}, {Eisenstein}, {Ferrer},
  {Maksimova}  \& {Pinto}}{{Garrison} et~al.}{2021}]{2021MNRAS.508..575G}
{Garrison} L.~H.,  {Eisenstein} D.~J.,  {Ferrer} D.,  {Maksimova} N.~A.,
  {Pinto} P.~A.,  2021, \mn@doi [\mnras] {10.1093/mnras/stab2482}, \href
  {https://ui.adsabs.harvard.edu/abs/2021MNRAS.508..575G} {508, 575}

\bibitem[\protect\citeauthoryear{{Genel} et~al.,}{{Genel}
  et~al.}{2014}]{2014MNRAS.445..175G}
{Genel} S.,  et~al., 2014, \mn@doi [\mnras] {10.1093/mnras/stu1654}, \href
  {https://ui.adsabs.harvard.edu/abs/2014MNRAS.445..175G} {445, 175}

\bibitem[\protect\citeauthoryear{{Givans} et~al.,}{{Givans}
  et~al.}{2022}]{2022JCAP...09..070G}
{Givans} J.~J.,  et~al., 2022, \mn@doi [\jcap] {10.1088/1475-7516/2022/09/070},
  \href {https://ui.adsabs.harvard.edu/abs/2022JCAP...09..070G} {2022, 070}

\bibitem[\protect\citeauthoryear{{Gnedin} \& {Hamilton}}{{Gnedin} \&
  {Hamilton}}{2002}]{Gnedin2002}
{Gnedin} N.~Y.,  {Hamilton} A. J.~S.,  2002, \mn@doi [\mnras]
  {10.1046/j.1365-8711.2002.05490.x}, \href
  {https://ui.adsabs.harvard.edu/abs/2002MNRAS.334..107G} {334, 107}

\bibitem[\protect\citeauthoryear{{Gouin}, {Gallo}  \& {Aghanim}}{{Gouin}
  et~al.}{2022}]{2022A&A...664A.198G}
{Gouin} C.,  {Gallo} S.,   {Aghanim} N.,  2022, \mn@doi [\aap]
  {10.1051/0004-6361/202243032}, \href
  {https://ui.adsabs.harvard.edu/abs/2022A&A...664A.198G} {664, A198}

\bibitem[\protect\citeauthoryear{{Gunn} \& {Peterson}}{{Gunn} \&
  {Peterson}}{1965}]{Gunn:1965ApJ...142.1633G}
{Gunn} J.~E.,  {Peterson} B.~A.,  1965, \mn@doi [\apj] {10.1086/148444}, \href
  {https://ui.adsabs.harvard.edu/abs/1965ApJ...142.1633G} {142, 1633}

\bibitem[\protect\citeauthoryear{{Hadzhiyska}, {Garrison}, {Eisenstein}  \&
  {Bose}}{{Hadzhiyska} et~al.}{2022}]{2022MNRAS.509.2194H}
{Hadzhiyska} B.,  {Garrison} L.~H.,  {Eisenstein} D.,   {Bose} S.,  2022,
  \mn@doi [\mnras] {10.1093/mnras/stab3066}, \href
  {https://ui.adsabs.harvard.edu/abs/2022MNRAS.509.2194H} {509, 2194}

\bibitem[\protect\citeauthoryear{{Hadzhiyska}, {Yuan}, {Blake}, {Garrison}  \&
  {Eisenstein}}{{Hadzhiyska} et~al.}{2023}]{Hadzhiyska+2023}
{Hadzhiyska} B.,  {Yuan} S.,  {Blake} C.,  {Garrison} L.,   {Eisenstein} D.,
  2023, submitted

\bibitem[\protect\citeauthoryear{{Hui} \& {Gnedin}}{{Hui} \&
  {Gnedin}}{1997}]{Hui:1997MNRAS.292...27H}
{Hui} L.,  {Gnedin} N.~Y.,  1997, \mn@doi [\mnras] {10.1093/mnras/292.1.27},
  \href {https://ui.adsabs.harvard.edu/abs/1997MNRAS.292...27H} {292, 27}

\bibitem[\protect\citeauthoryear{{Hui}, {Gnedin}  \& {Zhang}}{{Hui}
  et~al.}{1997}]{Hui:1997ApJ...486..599H}
{Hui} L.,  {Gnedin} N.~Y.,   {Zhang} Y.,  1997, \mn@doi [\apj]
  {10.1086/304539}, \href
  {https://ui.adsabs.harvard.edu/abs/1997ApJ...486..599H} {486, 599}

\bibitem[\protect\citeauthoryear{{Ir{\v{s}}i{\v{c}}}, {Viel}, {Haehnelt},
  {Bolton}  \& {Becker}}{{Ir{\v{s}}i{\v{c}}}
  et~al.}{2017a}]{2017PhRvL.119c1302I}
{Ir{\v{s}}i{\v{c}}} V.,  {Viel} M.,  {Haehnelt} M.~G.,  {Bolton} J.~S.,
  {Becker} G.~D.,  2017a, \mn@doi [\prl] {10.1103/PhysRevLett.119.031302},
  \href {https://ui.adsabs.harvard.edu/abs/2017PhRvL.119c1302I} {119, 031302}

\bibitem[\protect\citeauthoryear{{Ir{\v{s}}i{\v{c}}}
  et~al.,}{{Ir{\v{s}}i{\v{c}}} et~al.}{2017b}]{2017MNRAS.466.4332I}
{Ir{\v{s}}i{\v{c}}} V.,  et~al., 2017b, \mn@doi [\mnras]
  {10.1093/mnras/stw3372}, \href
  {https://ui.adsabs.harvard.edu/abs/2017MNRAS.466.4332I} {466, 4332}

\bibitem[\protect\citeauthoryear{{Kaiser}}{{Kaiser}}{1987}]{1987MNRAS.227....1K}
{Kaiser} N.,  1987, \mn@doi [\mnras] {10.1093/mnras/227.1.1}, \href
  {https://ui.adsabs.harvard.edu/abs/1987MNRAS.227....1K} {227, 1}

\bibitem[\protect\citeauthoryear{{Kirkby} et~al.,}{{Kirkby}
  et~al.}{2013}]{2013JCAP...03..024K}
{Kirkby} D.,  et~al., 2013, \mn@doi [\jcap] {10.1088/1475-7516/2013/03/024},
  \href {https://ui.adsabs.harvard.edu/abs/2013JCAP...03..024K} {2013, 024}

\bibitem[\protect\citeauthoryear{{LSST Dark Energy Science
  Collaboration}}{{LSST Dark Energy Science
  Collaboration}}{2012}]{2012arXiv1211.0310L}
{LSST Dark Energy Science Collaboration} 2012, arXiv e-prints, \href
  {https://ui.adsabs.harvard.edu/abs/2012arXiv1211.0310L} {p. arXiv:1211.0310}

\bibitem[\protect\citeauthoryear{{Le Goff} et~al.,}{{Le Goff}
  et~al.}{2011}]{LeGoff:2011A&A...534A.135L}
{Le Goff} J.~M.,  et~al., 2011, \mn@doi [\aap] {10.1051/0004-6361/201117736},
  \href {https://ui.adsabs.harvard.edu/abs/2011A&A...534A.135L} {534, A135}

\bibitem[\protect\citeauthoryear{{Levi} et~al.,}{{Levi}
  et~al.}{2013}]{2013arXiv1308.0847L}
{Levi} M.,  et~al., 2013, arXiv e-prints, \href
  {https://ui.adsabs.harvard.edu/abs/2013arXiv1308.0847L} {p. arXiv:1308.0847}

\bibitem[\protect\citeauthoryear{{Levi} et~al.,}{{Levi}
  et~al.}{2019}]{2019BAAS...51g..57L}
{Levi} M.,  et~al., 2019, in Bulletin of the American Astronomical Society.
  p.~57 (\mn@eprint {arXiv} {1907.10688})

\bibitem[\protect\citeauthoryear{{Maksimova}, {Garrison}, {Eisenstein},
  {Hadzhiyska}, {Bose}  \& {Satterthwaite}}{{Maksimova}
  et~al.}{2021}]{2021MNRAS.508.4017M}
{Maksimova} N.~A.,  {Garrison} L.~H.,  {Eisenstein} D.~J.,  {Hadzhiyska} B.,
  {Bose} S.,   {Satterthwaite} T.~P.,  2021, \mn@doi [\mnras]
  {10.1093/mnras/stab2484}, \href
  {https://ui.adsabs.harvard.edu/abs/2021MNRAS.508.4017M} {508, 4017}

\bibitem[\protect\citeauthoryear{{Marinacci} et~al.,}{{Marinacci}
  et~al.}{2018}]{2018MNRAS.480.5113M}
{Marinacci} F.,  et~al., 2018, \mn@doi [\mnras] {10.1093/mnras/sty2206}, \href
  {https://ui.adsabs.harvard.edu/abs/2018MNRAS.480.5113M} {480, 5113}

\bibitem[\protect\citeauthoryear{{McDonald} \& {Eisenstein}}{{McDonald} \&
  {Eisenstein}}{2007}]{2007PhRvD..76f3009M}
{McDonald} P.,  {Eisenstein} D.~J.,  2007, \mn@doi [\prd]
  {10.1103/PhysRevD.76.063009}, \href
  {https://ui.adsabs.harvard.edu/abs/2007PhRvD..76f3009M} {76, 063009}

\bibitem[\protect\citeauthoryear{{McDonald}, {Miralda-Escud{\'e}}, {Rauch},
  {Sargent}, {Barlow}, {Cen}  \& {Ostriker}}{{McDonald}
  et~al.}{2000}]{McDonald2000}
{McDonald} P.,  {Miralda-Escud{\'e}} J.,  {Rauch} M.,  {Sargent} W. L.~W.,
  {Barlow} T.~A.,  {Cen} R.,   {Ostriker} J.~P.,  2000, \mn@doi [\apj]
  {10.1086/317079}, \href
  {https://ui.adsabs.harvard.edu/abs/2000ApJ...543....1M} {543, 1}

\bibitem[\protect\citeauthoryear{{McDonald} et~al.,}{{McDonald}
  et~al.}{2005}]{McDonald2005}
{McDonald} P.,  et~al., 2005, \mn@doi [\apj] {10.1086/497563}, \href
  {https://ui.adsabs.harvard.edu/abs/2005ApJ...635..761M} {635, 761}

\bibitem[\protect\citeauthoryear{{McDonald} et~al.,}{{McDonald}
  et~al.}{2006}]{McDonald2006}
{McDonald} P.,  et~al., 2006, \mn@doi [\apjs] {10.1086/444361}, \href
  {https://ui.adsabs.harvard.edu/abs/2006ApJS..163...80M} {163, 80}

\bibitem[\protect\citeauthoryear{{Murgia}, {Ir{\v{s}}i{\v{c}}}  \&
  {Viel}}{{Murgia} et~al.}{2018}]{2018PhRvD..98h3540M}
{Murgia} R.,  {Ir{\v{s}}i{\v{c}}} V.,   {Viel} M.,  2018, \mn@doi [\prd]
  {10.1103/PhysRevD.98.083540}, \href
  {https://ui.adsabs.harvard.edu/abs/2018PhRvD..98h3540M} {98, 083540}

\bibitem[\protect\citeauthoryear{{Murgia}, {Scelfo}, {Viel}  \&
  {Raccanelli}}{{Murgia} et~al.}{2019}]{2019PhRvL.123g1102M}
{Murgia} R.,  {Scelfo} G.,  {Viel} M.,   {Raccanelli} A.,  2019, \mn@doi [\prl]
  {10.1103/PhysRevLett.123.071102}, \href
  {https://ui.adsabs.harvard.edu/abs/2019PhRvL.123g1102M} {123, 071102}

\bibitem[\protect\citeauthoryear{{Naiman} et~al.,}{{Naiman}
  et~al.}{2018}]{2018MNRAS.477.1206N}
{Naiman} J.~P.,  et~al., 2018, \mn@doi [\mnras] {10.1093/mnras/sty618}, \href
  {https://ui.adsabs.harvard.edu/abs/2018MNRAS.477.1206N} {477, 1206}

\bibitem[\protect\citeauthoryear{{Nelson} et~al.,}{{Nelson}
  et~al.}{2019}]{2019MNRAS.tmp.2010N}
{Nelson} D.,  et~al., 2019, \mn@doi [\mnras] {10.1093/mnras/stz2306}, \href
  {https://ui.adsabs.harvard.edu/abs/2019MNRAS.490.3234N} {490, 3234}

\bibitem[\protect\citeauthoryear{{Newman} et~al.,}{{Newman}
  et~al.}{2020}]{2020ApJ...891..147N}
{Newman} A.~B.,  et~al., 2020, \mn@doi [\apj] {10.3847/1538-4357/ab75ee}, \href
  {https://ui.adsabs.harvard.edu/abs/2020ApJ...891..147N} {891, 147}

\bibitem[\protect\citeauthoryear{{Nori}, {Murgia}, {Ir{\v{s}}i{\v{c}}}, {Baldi}
   \& {Viel}}{{Nori} et~al.}{2019}]{2019MNRAS.482.3227N}
{Nori} M.,  {Murgia} R.,  {Ir{\v{s}}i{\v{c}}} V.,  {Baldi} M.,   {Viel} M.,
  2019, \mn@doi [\mnras] {10.1093/mnras/sty2888}, \href
  {https://ui.adsabs.harvard.edu/abs/2019MNRAS.482.3227N} {482, 3227}

\bibitem[\protect\citeauthoryear{{Peebles} \& {Yu}}{{Peebles} \&
  {Yu}}{1970}]{Peebles:1970ApJ...162..815P}
{Peebles} P.~J.~E.,  {Yu} J.~T.,  1970, \mn@doi [\apj] {10.1086/150713}, \href
  {https://ui.adsabs.harvard.edu/abs/1970ApJ...162..815P} {162, 815}

\bibitem[\protect\citeauthoryear{{Peirani}, {Weinberg}, {Colombi}, {Blaizot},
  {Dubois}  \& {Pichon}}{{Peirani} et~al.}{2014a}]{2014ApJ...784...11P}
{Peirani} S.,  {Weinberg} D.~H.,  {Colombi} S.,  {Blaizot} J.,  {Dubois} Y.,
  {Pichon} C.,  2014a, \mn@doi [\apj] {10.1088/0004-637X/784/1/11}, \href
  {https://ui.adsabs.harvard.edu/abs/2014ApJ...784...11P} {784, 11}

\bibitem[\protect\citeauthoryear{{Peirani}, {Weinberg}, {Colombi}, {Blaizot},
  {Dubois}  \& {Pichon}}{{Peirani} et~al.}{2014b}]{Peirani:2014}
{Peirani} S.,  {Weinberg} D.~H.,  {Colombi} S.,  {Blaizot} J.,  {Dubois} Y.,
  {Pichon} C.,  2014b, \mn@doi [\apj] {10.1088/0004-637X/784/1/11}, \href
  {https://ui.adsabs.harvard.edu/abs/2014ApJ...784...11P} {784, 11}

\bibitem[\protect\citeauthoryear{{Peirani} et~al.,}{{Peirani}
  et~al.}{2022a}]{2022MNRAS.514.3222P}
{Peirani} S.,  et~al., 2022a, \mn@doi [\mnras] {10.1093/mnras/stac1344}, \href
  {https://ui.adsabs.harvard.edu/abs/2022MNRAS.514.3222P} {514, 3222}

\bibitem[\protect\citeauthoryear{{Peirani} et~al.,}{{Peirani}
  et~al.}{2022b}]{Peirani:2022}
{Peirani} S.,  et~al., 2022b, \mn@doi [\mnras] {10.1093/mnras/stac1344}, \href
  {https://ui.adsabs.harvard.edu/abs/2022MNRAS.514.3222P} {514, 3222}

\bibitem[\protect\citeauthoryear{{Perlmutter} et~al.,}{{Perlmutter}
  et~al.}{1999}]{Perlmutter:1999ApJ...517..565P}
{Perlmutter} S.,  et~al., 1999, \mn@doi [\apj] {10.1086/307221}, \href
  {https://ui.adsabs.harvard.edu/abs/1999ApJ...517..565P} {517, 565}

\bibitem[\protect\citeauthoryear{{Phillips}, {Weinberg}, {Croft}, {Hernquist},
  {Katz}  \& {Pettini}}{{Phillips} et~al.}{2001}]{Phillips2001}
{Phillips} J.,  {Weinberg} D.~H.,  {Croft} R. A.~C.,  {Hernquist} L.,  {Katz}
  N.,   {Pettini} M.,  2001, \mn@doi [\apj] {10.1086/322369}, \href
  {https://ui.adsabs.harvard.edu/abs/2001ApJ...560...15P} {560, 15}

\bibitem[\protect\citeauthoryear{{Pillepich} et~al.,}{{Pillepich}
  et~al.}{2018}]{2018MNRAS.473.4077P}
{Pillepich} A.,  et~al., 2018, \mn@doi [\mnras] {10.1093/mnras/stx2656}, \href
  {https://ui.adsabs.harvard.edu/abs/2018MNRAS.473.4077P} {473, 4077}

\bibitem[\protect\citeauthoryear{{Pillepich} et~al.,}{{Pillepich}
  et~al.}{2019}]{2019MNRAS.tmp.2024P}
{Pillepich} A.,  et~al., 2019, \mn@doi [\mnras] {10.1093/mnras/stz2338}, \href
  {https://ui.adsabs.harvard.edu/abs/2019MNRAS.490.3196P} {490, 3196}

\bibitem[\protect\citeauthoryear{{Qezlou}, {Newman}, {Rudie}  \&
  {Bird}}{{Qezlou} et~al.}{2022}]{2022ApJ...930..109Q}
{Qezlou} M.,  {Newman} A.~B.,  {Rudie} G.~C.,   {Bird} S.,  2022, \mn@doi
  [\apj] {10.3847/1538-4357/ac6259}, \href
  {https://ui.adsabs.harvard.edu/abs/2022ApJ...930..109Q} {930, 109}

\bibitem[\protect\citeauthoryear{{Riess} et~al.,}{{Riess}
  et~al.}{1998}]{Riess:1998AJ....116.1009R}
{Riess} A.~G.,  et~al., 1998, \mn@doi [\aj] {10.1086/300499}, \href
  {https://ui.adsabs.harvard.edu/abs/1998AJ....116.1009R} {116, 1009}

\bibitem[\protect\citeauthoryear{{Rogers} \& {Peiris}}{{Rogers} \&
  {Peiris}}{2021a}]{2021PhRvD.103d3526R}
{Rogers} K.~K.,  {Peiris} H.~V.,  2021a, \mn@doi [\prd]
  {10.1103/PhysRevD.103.043526}, \href
  {https://ui.adsabs.harvard.edu/abs/2021PhRvD.103d3526R} {103, 043526}

\bibitem[\protect\citeauthoryear{{Rogers} \& {Peiris}}{{Rogers} \&
  {Peiris}}{2021b}]{2021PhRvL.126g1302R}
{Rogers} K.~K.,  {Peiris} H.~V.,  2021b, \mn@doi [\prl]
  {10.1103/PhysRevLett.126.071302}, \href
  {https://ui.adsabs.harvard.edu/abs/2021PhRvL.126g1302R} {126, 071302}

\bibitem[\protect\citeauthoryear{{Seljak} et~al.,}{{Seljak}
  et~al.}{2005}]{Seljak2005}
{Seljak} U.,  et~al., 2005, \mn@doi [\prd] {10.1103/PhysRevD.71.103515}, \href
  {https://ui.adsabs.harvard.edu/abs/2005PhRvD..71j3515S} {71, 103515}

\bibitem[\protect\citeauthoryear{{Seljak}, {Slosar}  \& {McDonald}}{{Seljak}
  et~al.}{2006}]{Seljak2006}
{Seljak} U.,  {Slosar} A.,   {McDonald} P.,  2006, \mn@doi [\jcap]
  {10.1088/1475-7516/2006/10/014}, \href
  {https://ui.adsabs.harvard.edu/abs/2006JCAP...10..014S} {2006, 014}

\bibitem[\protect\citeauthoryear{{Silber} et~al.,}{{Silber}
  et~al.}{2022}]{2022arXiv220509014S}
{Silber} J.~H.,  et~al., 2022, arXiv e-prints, \href
  {https://ui.adsabs.harvard.edu/abs/2022arXiv220509014S} {p. arXiv:2205.09014}

\bibitem[\protect\citeauthoryear{{Sinigaglia}, {Kitaura},
  {Balaguera-Antol{\'\i}nez}, {Shimizu}, {Nagamine}, {S{\'a}nchez-Benavente}
  \& {Ata}}{{Sinigaglia} et~al.}{2022}]{2022ApJ...927..230S}
{Sinigaglia} F.,  {Kitaura} F.-S.,  {Balaguera-Antol{\'\i}nez} A.,  {Shimizu}
  I.,  {Nagamine} K.,  {S{\'a}nchez-Benavente} M.,   {Ata} M.,  2022, \mn@doi
  [\apj] {10.3847/1538-4357/ac5112}, \href
  {https://ui.adsabs.harvard.edu/abs/2022ApJ...927..230S} {927, 230}

\bibitem[\protect\citeauthoryear{{Sorini}, {O{\~n}orbe}, {Luki{\'c}}  \&
  {Hennawi}}{{Sorini} et~al.}{2016a}]{2016ApJ...827...97S}
{Sorini} D.,  {O{\~n}orbe} J.,  {Luki{\'c}} Z.,   {Hennawi} J.~F.,  2016a,
  \mn@doi [\apj] {10.3847/0004-637X/827/2/97}, \href
  {https://ui.adsabs.harvard.edu/abs/2016ApJ...827...97S} {827, 97}

\bibitem[\protect\citeauthoryear{{Sorini}, {O{\~n}orbe}, {Luki{\'c}}  \&
  {Hennawi}}{{Sorini} et~al.}{2016b}]{Sorini:2016}
{Sorini} D.,  {O{\~n}orbe} J.,  {Luki{\'c}} Z.,   {Hennawi} J.~F.,  2016b,
  \mn@doi [\apj] {10.3847/0004-637X/827/2/97}, \href
  {https://ui.adsabs.harvard.edu/abs/2016ApJ...827...97S} {827, 97}

\bibitem[\protect\citeauthoryear{{Sorini}, {O{\~n}orbe}, {Hennawi}  \&
  {Luki{\'c}}}{{Sorini} et~al.}{2018}]{2018ApJ...859..125S}
{Sorini} D.,  {O{\~n}orbe} J.,  {Hennawi} J.~F.,   {Luki{\'c}} Z.,  2018,
  \mn@doi [\apj] {10.3847/1538-4357/aabb52}, \href
  {https://ui.adsabs.harvard.edu/abs/2018ApJ...859..125S} {859, 125}

\bibitem[\protect\citeauthoryear{{Spergel} et~al.,}{{Spergel}
  et~al.}{2003}]{Spergel2003}
{Spergel} D.~N.,  et~al., 2003, \mn@doi [\apjs] {10.1086/377226}, \href
  {https://ui.adsabs.harvard.edu/abs/2003ApJS..148..175S} {148, 175}

\bibitem[\protect\citeauthoryear{{Springel}}{{Springel}}{2010}]{2010MNRAS.401..791S}
{Springel} V.,  2010, \mn@doi [\mnras] {10.1111/j.1365-2966.2009.15715.x},
  \href {https://ui.adsabs.harvard.edu/abs/2010MNRAS.401..791S} {401, 791}

\bibitem[\protect\citeauthoryear{{Springel} et~al.,}{{Springel}
  et~al.}{2018}]{2018MNRAS.475..676S}
{Springel} V.,  et~al., 2018, \mn@doi [\mnras] {10.1093/mnras/stx3304}, \href
  {https://ui.adsabs.harvard.edu/abs/2018MNRAS.475..676S} {475, 676}

\bibitem[\protect\citeauthoryear{{Stark}, {Font-Ribera}, {White}  \&
  {Lee}}{{Stark} et~al.}{2015}]{2015MNRAS.453.4311S}
{Stark} C.~W.,  {Font-Ribera} A.,  {White} M.,   {Lee} K.-G.,  2015, \mn@doi
  [\mnras] {10.1093/mnras/stv1868}, \href
  {https://ui.adsabs.harvard.edu/abs/2015MNRAS.453.4311S} {453, 4311}

\bibitem[\protect\citeauthoryear{{Tassev}, {Zaldarriaga}  \&
  {Eisenstein}}{{Tassev} et~al.}{2013}]{Tassev:2013JCAP...06..036T}
{Tassev} S.,  {Zaldarriaga} M.,   {Eisenstein} D.~J.,  2013, \mn@doi [\jcap]
  {10.1088/1475-7516/2013/06/036}, \href
  {https://ui.adsabs.harvard.edu/abs/2013JCAP...06..036T} {2013, 036}

\bibitem[\protect\citeauthoryear{{Verde} et~al.,}{{Verde}
  et~al.}{2003}]{Verde2003}
{Verde} L.,  et~al., 2003, \mn@doi [\apjs] {10.1086/377335}, \href
  {https://ui.adsabs.harvard.edu/abs/2003ApJS..148..195V} {148, 195}

\bibitem[\protect\citeauthoryear{{Viel} \& {Haehnelt}}{{Viel} \&
  {Haehnelt}}{2006}]{2006MNRAS.365..231V}
{Viel} M.,  {Haehnelt} M.~G.,  2006, \mn@doi [\mnras]
  {10.1111/j.1365-2966.2005.09703.x}, \href
  {https://ui.adsabs.harvard.edu/abs/2006MNRAS.365..231V} {365, 231}

\bibitem[\protect\citeauthoryear{{Viel}, {Haehnelt}  \& {Springel}}{{Viel}
  et~al.}{2004a}]{2004MNRAS.354..684V}
{Viel} M.,  {Haehnelt} M.~G.,   {Springel} V.,  2004a, \mn@doi [\mnras]
  {10.1111/j.1365-2966.2004.08224.x}, \href
  {https://ui.adsabs.harvard.edu/abs/2004MNRAS.354..684V} {354, 684}

\bibitem[\protect\citeauthoryear{{Viel}, {Weller}  \& {Haehnelt}}{{Viel}
  et~al.}{2004b}]{Viel2004b}
{Viel} M.,  {Weller} J.,   {Haehnelt} M.~G.,  2004b, \mn@doi [\mnras]
  {10.1111/j.1365-2966.2004.08498.x}, \href
  {https://ui.adsabs.harvard.edu/abs/2004MNRAS.355L..23V} {355, L23}

\bibitem[\protect\citeauthoryear{{Vogelsberger} et~al.,}{{Vogelsberger}
  et~al.}{2014a}]{2014MNRAS.444.1518V}
{Vogelsberger} M.,  et~al., 2014a, \mn@doi [\mnras] {10.1093/mnras/stu1536},
  \href {https://ui.adsabs.harvard.edu/abs/2014MNRAS.444.1518V} {444, 1518}

\bibitem[\protect\citeauthoryear{{Vogelsberger} et~al.,}{{Vogelsberger}
  et~al.}{2014b}]{2014Natur.509..177V}
{Vogelsberger} M.,  et~al., 2014b, \mn@doi [\nat] {10.1038/nature13316}, \href
  {https://ui.adsabs.harvard.edu/abs/2014Natur.509..177V} {509, 177}

\bibitem[\protect\citeauthoryear{{Weinberger} et~al.,}{{Weinberger}
  et~al.}{2017}]{2017MNRAS.465.3291W}
{Weinberger} R.,  et~al., 2017, \mn@doi [\mnras] {10.1093/mnras/stw2944}, \href
  {https://ui.adsabs.harvard.edu/abs/2017MNRAS.465.3291W} {465, 3291}

\bibitem[\protect\citeauthoryear{{Weinberger}, {Springel}  \&
  {Pakmor}}{{Weinberger} et~al.}{2020}]{2019arXiv190904667W}
{Weinberger} R.,  {Springel} V.,   {Pakmor} R.,  2020, \mn@doi [\apjs]
  {10.3847/1538-4365/ab908c}, \href
  {https://ui.adsabs.harvard.edu/abs/2020ApJS..248...32W} {248, 32}

\bibitem[\protect\citeauthoryear{{Y{\`e}che}, {Palanque-Delabrouille}, {Baur}
  \& {du Mas des Bourboux}}{{Y{\`e}che} et~al.}{2017}]{2017JCAP...06..047Y}
{Y{\`e}che} C.,  {Palanque-Delabrouille} N.,  {Baur} J.,   {du Mas des
  Bourboux} H.,  2017, \mn@doi [\jcap] {10.1088/1475-7516/2017/06/047}, \href
  {https://ui.adsabs.harvard.edu/abs/2017JCAP...06..047Y} {2017, 047}

\bibitem[\protect\citeauthoryear{{York} et~al.,}{{York}
  et~al.}{2000}]{2000AJ....120.1579Y}
{York} D.~G.,  et~al., 2000, \mn@doi [\aj] {10.1086/301513}, \href
  {https://ui.adsabs.harvard.edu/abs/2000AJ....120.1579Y} {120, 1579}

\bibitem[\protect\citeauthoryear{{Yuan}, {Garrison}, {Hadzhiyska}, {Bose}  \&
  {Eisenstein}}{{Yuan} et~al.}{2022}]{2022MNRAS.510.3301Y}
{Yuan} S.,  {Garrison} L.~H.,  {Hadzhiyska} B.,  {Bose} S.,   {Eisenstein}
  D.~J.,  2022, \mn@doi [\mnras] {10.1093/mnras/stab3355}, \href
  {https://ui.adsabs.harvard.edu/abs/2022MNRAS.510.3301Y} {510, 3301}

\bibitem[\protect\citeauthoryear{{Zaldarriaga}, {Hui}  \&
  {Tegmark}}{{Zaldarriaga} et~al.}{2001}]{2001ApJ...557..519Z}
{Zaldarriaga} M.,  {Hui} L.,   {Tegmark} M.,  2001, \mn@doi [\apj]
  {10.1086/321652}, \href
  {https://ui.adsabs.harvard.edu/abs/2001ApJ...557..519Z} {557, 519}

\bibitem[\protect\citeauthoryear{{du Mas des Bourboux} et~al.,}{{du Mas des
  Bourboux} et~al.}{2020}]{2020ApJ...901..153D}
{du Mas des Bourboux} H.,  et~al., 2020, \mn@doi [\apj]
  {10.3847/1538-4357/abb085}, \href
  {https://ui.adsabs.harvard.edu/abs/2020ApJ...901..153D} {901, 153}

\makeatother
\end{thebibliography}

%\begin{thebibliography}{99}
%\end{thebibliography}

%%%%%%%%%%%%%%%%%%%%%%%%%%%%%%%%%%%%%%%%%%%%%%%%%%

%%%%%%%%%%%%%%%%% APPENDICES %%%%%%%%%%%%%%%%%%%%%

\appendix

\section{Convolution with the Doppler profile}
\label{sec:voigt}

When applying redshift space distortions in our FGPA-based models, we ignore the effects of thermal broadening due to the random thermal velocities of the gas atoms. In this Appendix, we illustrate the impact thermal broadening, as implemented through a Doppler profile convolution, has on our measured 1D power spectrum. {The obvious advantage of the Voigt profile is that it incorporates a physical effect and thus adds more realism to the very small-scale behavior. On the other hand, it is computationally more expensive than the alternative and ends up yielding qualitatively similar results to what we obtain when we add small-scale noise (see Section~\ref{sec:noise}). Below, we will show that adopting the Voigt profile has an almost negligible effect on the power spectrum given the current level of accuracy of our mocks and scales of interest for the power spectrum, $k \lesssim 3 \hompc$, and thus, we can justify omit it from the present mocks.}

To obtain the Doppler-profile-convolved optical depth in redshift space, we perform the integral over velocity space for each skewer:
\begin{equation}
    \tau(s) = \int dx \frac{\tau(x)}{b(x)} \exp{\left[-\left(\frac{ (s - x - v_r(x))}{b(x)}\right)^2 \right]} ,
    \label{eq:voigt}
\end{equation}
where $s$ and $x$ are velocity coordinates, $v_r$ is the peculiar velocity in the line-of-sight direction, and 
\begin{equation}
    b(x) \equiv \sqrt{\frac{2 k_B T(x)}{m_p}} 
\end{equation}
is the thermal velocity of the atoms, $k_B$ the Boltzmann constant, and $m_p$ the proton mass. We approximate the temperature of the gas as:
\begin{equation}
    T(x) = T_0 \left[1 + \delta_{\rm dm}(x)\right]^{\gamma - 1 }
\end{equation}
with $\delta_{\rm dm}$ being the dark-matter overdensity obtained via TSC interpolation and $T_0 = 1.94 \times 10^4 \ {\rm K}$ being the normalization factor \citep{2022ApJ...930..109Q}.

\begin{figure}
    \centering
    \includegraphics[width=0.48\textwidth]{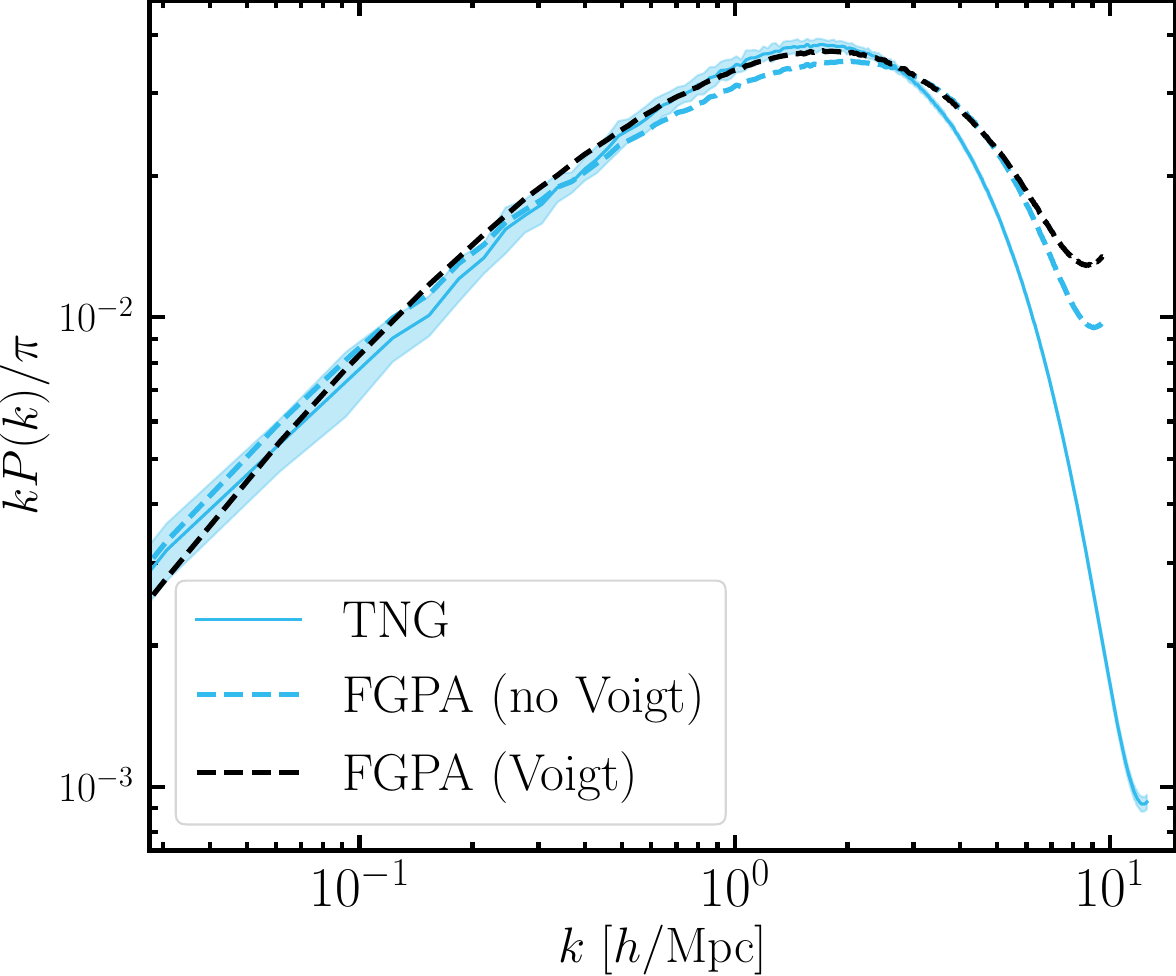}
    \caption{1D power spectrum comparison between the ``true'' \lyaf\ extracted from TNG300-1 and the FGPA-generated skewers for the fiducial model (blue solid; see Table~\ref{tab:models}) with and without a convolution with the Doppler profile (black and blue dashed, respectively; see Eq.~\ref{eq:voigt}). Note that we refit $\tau_0$ and $\sigma_\epsilon$ (see Section~\ref{sec:mocks}) to fit the mean and variance of the flux.} 
    \label{fig:voigt}
\end{figure}

From Fig.~\ref{fig:voigt}, we see that indeed the effect of convolving with the Doppler profile on the 1D power spectrum is negligible for our model, although it does appear to boost slightly the power near $k \sim 1 \hompc$, reducing the discrepancy between the two curves. We note that after applying the convolution, we refit $\tau_0$ and $\sigma_\epsilon$ (see Section~\ref{sec:mocks}) to fit the mean and variance of the flux. Additionally, we expect thermal broadening to lead to a suppression of the power on scales smaller than $k > 1 \hompc$, which would add more realism to our mocks and lead to a better agreement with the hydro simulation on these scales (see the top panels of Fig.~\ref{fig:power_abacus}). We plan to incorporate this effect into future versions of our mocks, {as the change is negligible given our current precision and scales of interest.} %We note that the affected scales are typically smaller than the resolution of our mocks and hence, we do not expect to see sizeable differences.

%\section{Effect of FGPA model on BAO peak broadening}
%\label{sec:bao}

%%%%%%%%%%%%%%%%%%%%%%%%%%%%%%%%%%%%%%%%%%%%%%%%%%

% Don't change these lines
\bsp	% typesetting comment
\label{lastpage}
\end{document}